\documentclass[a4paper,fleqn]{cas-dc}
\usepackage{lmodern}
\usepackage{fix-cm} 
\fontsize{10pt}{12pt}\selectfont
\usepackage{inconsolata}
\usepackage{makecell}
\usepackage{newtxtext}

\usepackage[authoryear,longnamesfirst]{natbib}
\usepackage[linesnumbered,ruled,vlined]{algorithm2e}
\usepackage{lipsum} 
\usepackage{nomencl} 
\usepackage{caption}

\usepackage{tabularx} 
\usepackage{booktabs} 
\usepackage{algorithm2e}
\usepackage[linesnumbered,ruled,vlined]{algorithm2e}

\usepackage{array}
\usepackage{tikz}
\usepackage{amssymb}
\usepackage[utf8]{inputenc}
\usepackage{latexsym}
\usepackage{amsfonts}
\usepackage{amsmath}
\usepackage{graphicx}
\usepackage{algpseudocode}
\usepackage{float}
\usepackage[english]{babel}
\usepackage{anyfontsize}
\usepackage{geometry}
\usepackage{hyperref}
\usepackage{enumitem}
\usepackage{xcolor}
\usepackage{caption} 
\def\tsc#1{\csdef{#1}{\textsc{\lowercase{#1}}\xspace}}
\tsc{WGM}
\tsc{QE}
\tsc{EP}
\tsc{PMS}
\tsc{BEC}
\tsc{DE}

\begin{document}
\let\WriteBookmarks\relax
\def\floatpagepagefraction{1}
\def\textpagefraction{.001}

\shorttitle{}

\shortauthors{Varshney et~al.}

\title [mode = title]{ A Login Page Transparency and Visual Similarity-Based Zero-Day Phishing Defense Protocol}                      
\tnotemark[1,2]

\tnotetext[1]{This document is the results of the research
   project funded by the National Science Foundation.}

\tnotetext[2]{The second title footnote which is a longer text matter
   to fill through the whole text width and overflow into
   another line in the footnotes area of the first page.}

%
\author[1]{Gaurav Varshney}[type=editor,
                        auid=000,bioid=1,
                        orcid=0000-0001-7511-2910]



\ead{gaurav.varshney@iitjammu.ac.in}

\ead[url]{https://www.iitjammu.ac.in/faculty/~gauravvarshney}

\credit{Conceptualization of this study, Methodology, Software}

\affiliation[1]{organization={Indian Institute of Technology Jammu},
    addressline={J\&K }, 
    postcode={181221}, 
    country={India}}

\author[1]{Akanksha Raj}[style=chinese]
\ead{2020ucs0109@iitjammu.ac.in}

\author[1]{Divya Sangwan}[%
   ]
\ead{2020ucs0093@iitjammu.ac.in}

\credit{Data curation, Writing - Original draft preparation}

\affiliation[2]{organization={CSIRO's Data61, Australia},
    city={Level 5/13 Garden St},
    state={ Eveleigh NSW 2015},
    country={Australia}}

\author%
[2]
{Sharif Abuadbba}
\ead{ data61.csiro.au}

\affiliation[3]{organization={University of Western Australia},
    city={Perth},
    postcode={6009}, 
    country={Australia}}
\author%
[1]
{Rina Mishra}
\cormark[1]
\ead{rina.mishra@iitjammu.ac.in}
\author%
[3]
{Yansong Gao}
\ead{garrison.gao@uwa.edu.au}
\cortext[cor1]{Corresponding author}
\cortext[cor2]{Principal corresponding author}

\fntext[fn1]{This is the first author footnote. but is common to third
  author as well.}
\fntext[fn2]{Another author footnote, this is a very long footnote and
  it should be a really long footnote. But this footnote is not yet
  sufficiently long enough to make two lines of footnote text.}

\nonumnote{This note has no numbers. In this work we demonstrate $a_b$
  the formation Y\_1 of a new type of polariton on the interface
  between a cuprous oxide slab and a polystyrene micro-sphere placed
  on the slab.
  }
\begin{abstract}
Phishing is a prevalent cyberattack that uses look-alike websites to deceive users into revealing sensitive information. Numerous efforts have been made by the Internet community and security organizations to detect, prevent, or train users to avoid falling victim to phishing attacks. Most of this research over the years has been highly diverse and application-oriented, often serving as standalone solutions for HTTP clients, servers, or third parties. However, limited work has been done to develop a comprehensive or proactive protocol-oriented solution to effectively counter phishing attacks. Inspired by the concept of certificate transparency, which allows certificates issued by Certificate Authorities (CAs) to be publicly verified by clients, thereby enhancing transparency, we propose a concept called Page Transparency (PT) for the web. The proposed PT requires login pages that capture users' sensitive information to be publicly logged via PLS and made available to web clients for verification. The pages are verified to be logged using cryptographic proofs. Since all pages are logged on a PLS and visually compared with existing pages through a comprehensive visual page-matching algorithm, it becomes impossible for an attacker to register a deceptive look-alike page on the PLS and receive the cryptographic proof required for client verification. All implementations occur on the client side, facilitated by the introduction of a new HTTP PT header, eliminating the need for platform-specific changes or the installation of third-party solutions for phishing prevention. 
\end{abstract}



\maketitle
\begin{keywords}
Certificate Transparency(CT) \sep Domain Owner Server(DOS) \sep Public Log Server(PLS) \sep Siamese Network Model(SNM) \sep Signed Page Timestamp(SPT) \sep Phishing Detection \sep Phishing Prevention \sep Deep Learning \sep User Authentication \sep Cybersecurity
\end{keywords}

\section{Introduction}
Phishing incidents have a long history, dating back to 1996 when the first phishing attack was carried out by impersonating AOL (America Online) employees \cite{13}. Since then, the number of phishing attacks has steadily increased despite the availability of numerous anti-phishing solutions. In the first quarter of 2024, the APWG report recorded 963,994 phishing attacks \cite{3}, representing a 46.93\% increase in newly identified attacks from 2022 to 2023.

Many solutions to combat phishing incidents have been proposed, including Phishing Detection and Phishing Prevention. Phishing detection solutions implemented so far include list-based approaches, where the URL being visited is checked against a whitelist \cite{6,7,40,41,42,43,44,45,48} (which maintains a list of trusted domains). If the URL is not found there, it is then checked against a blacklist (which maintains a list of blocked URLs) \cite{8,9,46,47,49,50,52,53,54,55,56,57}. However, these solutions require continuous list updates and typically take at least 24 hours to categorize a URL into a specific list.

Next, URL feature-based solutions were introduced, where the current URL is compared against a set of features characteristic of malicious URLs to identify phishing attempts \cite{4,5}. To handle real-time, zero-day phishing threats, search engine-based solutions \cite{14} were developed, which query search engines using a combination of elements such as the title, domain, copyright information, favicon, keywords, and logos to identify phishing sites \cite{11}. Later, heuristic and machine learning (ML)-based solutions were introduced, which extract features from HTML content, DOM structure, and the URL \cite{38,39} itself to train ML models. Phishing detection is based on the model's output. However, once these techniques are understood, attackers can easily bypass them.

In visual similarity-based solutions \cite{2,15}, a current snapshot of the webpage is compared against a dataset of webpage screenshots, logos, or favicons to assess the visual similarity of the visited site. While this technique delivers accurate results for websites within the dataset, it requires constant, resource-intensive dataset updates. Several approaches also combine different methods \cite{61}, such as integrating search engine-based \cite{59,60} and URL-based phishing detection\cite{58}. However, all of these solutions are limited to the records available in their respective datasets. Thus, no phishing detection method has proven foolproof.

\begin{figure*}
    \centering
    \includegraphics[width=0.8\textwidth]{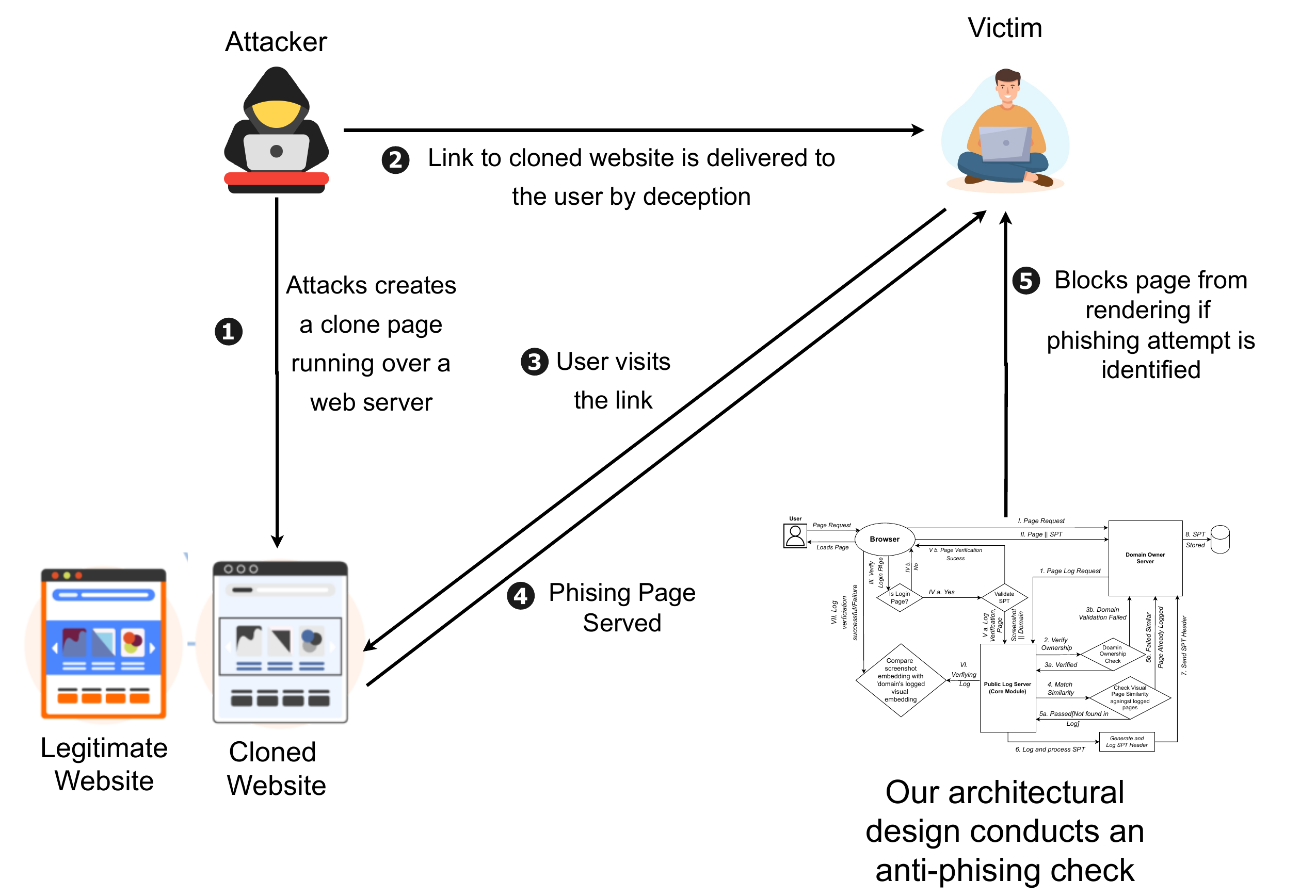}
    \caption{\label{fig:pal-1}Phishing Attack Lifecycle}
\end{figure*}

\textbf{Phishing prevention} solutions have also been developed, including two-factor authentication, biometric authentication, HTTPS/TLS-based solutions \cite{1}, and certificate transparency-based solutions \cite{16}, which maintain a public log server(PLS) to record all certificates issued by certificate authorities (CAs). Drawing inspiration from Certificate Transparency, we propose the Login Page Transparency protocol, which not only detects phishing but also prevents page rendering if the page does not meet security standards.  Figure 1 illustrates the lifecycle of a phishing attack and highlights the role of our detection and protection mechanism within the lifecycle. The attacker targets a legitimate website, creating a cloned version to execute the phishing attack. When a user attempts to visit the legitimate site, the cloned page is served instead. This is where our solution intervenes, performing visual similarity checks and preventing the phishing page from rendering on the user's screen.

\noindent The key contributions of the proposed method include: 

\begin{itemize} 
\item To the best of our knowledge, this is the first paper to introduce Page Transparency as a solution for both phishing detection and prevention. 
\item The novelty of the proposed method lies in the concept of Page Transparency, which halts page rendering if the page is not logged into the PLS. 
\item We propose a unique dataset consisting solely of credential-requiring pages/login pages, targeting 91 distinct brands, including 2,325 phishing pages, 115 legitimate pages, and 50 unseen legitimate web pages. 
\end{itemize}
\section{Related Works}
The biggest challenge in Internet security is ensuring that web users are communicating with the legitimate website corresponding to a domain. To address these challenges, several solutions have been introduced.
The web request process from the client to the server begins with translating a human-readable host name into a machine-readable IP address, a task handled by DNS. However, DNS is an open resolver and is therefore vulnerable to various attacks. To protect DNS cache poisoning leading to phishing attacks and bring a level of data origin authentication and integrity, the IETF introduced DNSSEC (Domain Name System Security Extensions), but DNSSEC is not widely being used and several servers are still running classic DNS and are still susceptible to phishing, spoofing, and man-in-the-middle (MITM) attacks \cite{17,18}. Although combining DNSSEC with an SSL certificate issued by a Certificate Authority (CA) can provide some protection, there are too many CAs, making it difficult to trust all of them. A compromised CA can issue a valid certificate for any domain name, thus presenting a major security concern. To mitigate this, DANE (DNS-Based Authentication of Named Entities) was introduced, limiting the number of CAs that can be trusted by a client and preventing any untrusted CA from issuing certificates for a domain. CAA was a similar policy that can be kept in DNS to announce trusted CAs that can issue a TLS certificate for a domain. Essentially, DANE \cite{18} is an internet security protocol that leverages the DNS hierarchy and allows the domain owners to publish their certificates integrity values in TLSA RR to be verified whenever someone establishes a connection provide a second level of trust on the TLS certificates.

However, a drawback of this approach is that the TLSA record (used in DANE) is simply a hash value. If a malicious actor creates a Public Key Infrastructure (PKI) hierarchy with a self-signed certificate and issues fake certificates to end entities, browsers will not trust these certificates, but DANE would still validate their TLSA records as legitimate.

To address this issue, \cite{17} proposed the DANE-based Trusted Server, which acts as a third-party validator for the certificates of all entities within the network. However, this solution faces its own challenge: if the Trusted Server itself is compromised, it could undermine the security of the entire system. 

    Additionally, \cite{1} proposed a detection method that focuses on the footprints of TLS certificates to address the aforementioned challenge. By collecting TLS certificates of phishing websites from Certificate Transparency logs and thoroughly analyzing them, this method enables monitoring of all HTTPS websites. It proves effective even when attackers use dynamic DNS (DDNS) or hosting services, which might bypass traditional domain registration methods. This approach helps detect phishing websites before they are published on the Internet. However, this method has limitations, including its reliance on URLs collected from the OpenPhish dataset, which could introduce bias. Additionally, it struggles with phishing detection involving wildcard certificates, as these can obscure the detection process.

Recently, \cite{19} proposed the DoH (DNS over HTTPS) protocol to ship DNS request responses as HTTP payload from browsers to increase trust and provide confidentiality and integrity services. A trusted DoH server can handle both domain name resolution and public key authentication, thereby reducing dependency on CAs. In this protocol, the web server sends a Certificate Signing Request (CSR) for authentication along with a Time to Live (TTL) for the public key. DoH then generates a random challenge and sends it to the web server to prove domain ownership. The web server hosts the solution on a link known to the DoH server, which verifies the presence of the verification token by visiting the URL. If the DoH server finds the file reachable, it stores the server’s information. When a client wants to connect to a web server, it sends an access request to the DoH server, which simultaneously handles domain name resolution and public key authentication.

While DoH servers can handle substantial traffic, the added responsibility of public key authentication could strain their resources, leading to delays or potential service disruptions. This approach centralizes trust from multiple CAs to a smaller number of DoH providers. While these providers (e.g., Google, Cloudflare) are generally reliable, centralizing trust in fewer entities could create single points of failure and increase the risk of large-scale attacks if these providers are compromised. 

   The drawbacks of the aforementioned solutions highlight the necessity for Certificate Transparency (CT) \cite{16}. CT is a public, verifiable, append-only log that records all issued certificates. It is cryptographically verifiable, meaning that clients can check whether certificates are logged, and servers can monitor logs for issued certificates. CT simplifies decision-making by not relying on users: the certificate is either logged or it is not. If it is logged, the corresponding server operator can take action if necessary. If it is not logged, the browser can automatically decline the connection. The availability of a certificate in log provide transparency of the subject and issuer of the certificate to all public stakeholders. 

To comply with these Certificate Transparency solutions, \cite{20} developed Phish-Hook. Phish-Hook employs machine learning algorithms to classify the likelihood that a certificate is associated with a phishing website. The model uses eight specific features extracted from CT log data. However, this solution suffers from the dataset imbalance problem, and its effectiveness heavily relies on the completeness and accuracy of CT logs. Additionally, when encountering unauthenticated websites, the system falls back to the traditional PKI method, reintroducing the same vulnerabilities Phish-Hook aims to mitigate.

We identified that CT is a very promising solution for detecting certificates issued by rouge CAs or when the CAs go malicious and issue certificates or a domain to attackers. We see that the problem of Anti-Phishing is similar. By closely monitoring the similarities of the problem statement we identified that phishing pages are also issued/created arbitrarily by attackers on the internet which are most of the times happen to be login pages or important pages of site where the attacker is interested in stealing form data. The pages if considered to be certificates and phishers as CA one can use the concept of login page transparency by logging all login pages of a website in a publicly verifiable log to see if an attacker has also not created a similar page on the Internet. In an analogous way to CT the LPT can be managed by web clients such as browsers where the browsers can receive pages with LPT logs to be verified before opening the pages and browsers can open only those pages logged in any of the PLS. After a rigorous study of 2 years, we have evolved a concept of page transparency which is motivated from the concept of CT in this paper that proactively detects and prevents phishing over the Internet.

\section{Proposed Work: Visual Transparency for Anti-phishing}
In this paper, we present an innovative approach to anti-phishing by introducing an architectural change in web page access that proactively thwarts phishing attacks. Our solution is inspired by the well-established architecture of certificate transparency. We propose a new architecture that introduces login-page transparency with the use of public login page logs. This multiphase architecture integrates a PLS, a  Page Transparency-enabled Browser, a custom SPT Header, and the Domain Owner to create a system that addresses phishing attacks at their core.



\begin{figure*}
    \centering
    \includegraphics[width=1\textwidth, height=20cm, keepaspectratio, trim=20pt 1300pt 10pt 20pt, clip]{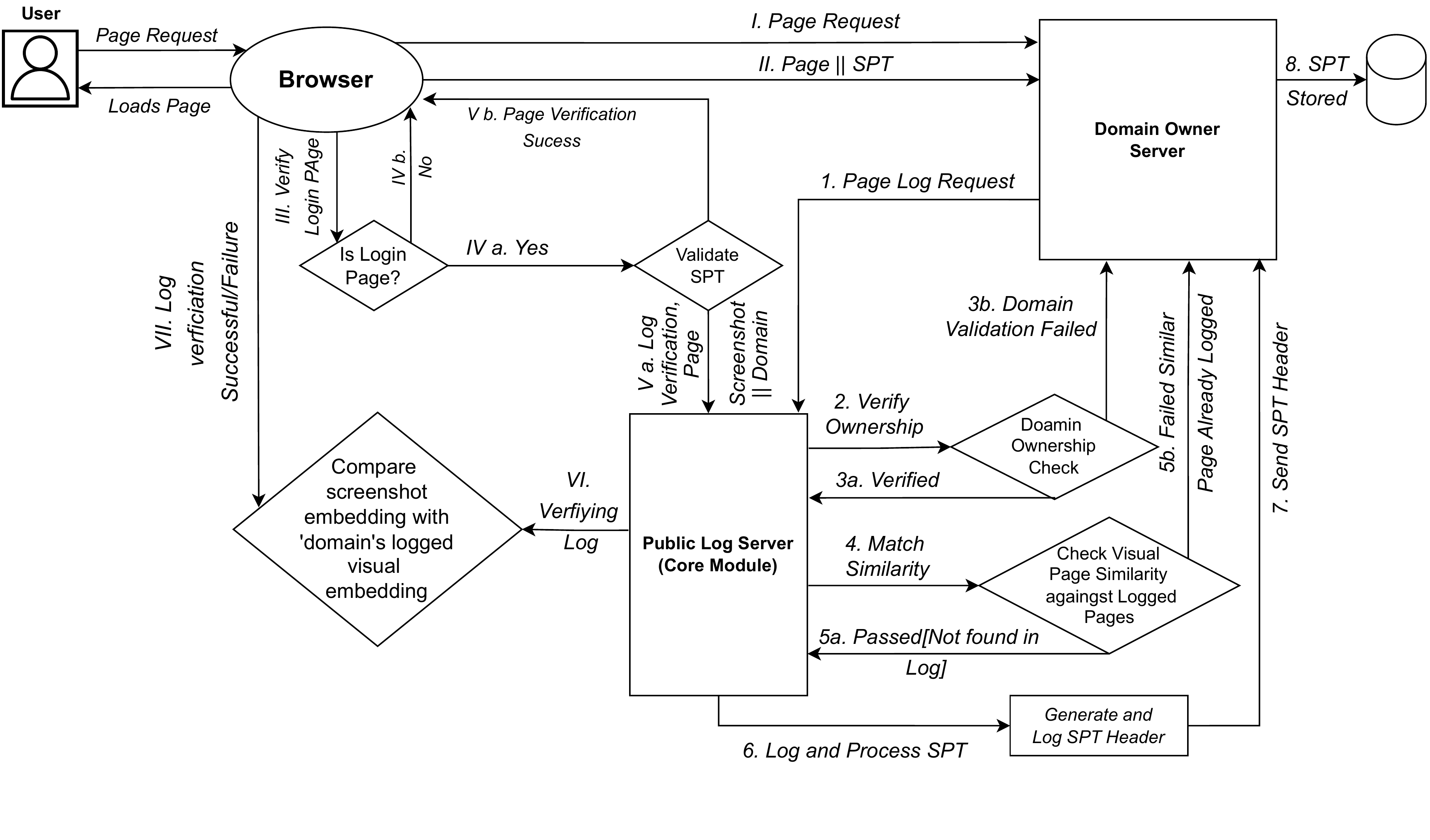}
    \caption{Overall Architecture}
    \label{fig:architecture}
\end{figure*}

In the proposed architecture, every domain owner requests PLS for SPT (Signed Page Timestamp) for its login page. PLS checks the visual similarity of this login page to the stored embeddings of existing logged pages of other domains to prevent a phisher logging a target domain's look-alike login page that has been already logged via the original domain. This is a one-time process. If the page is not visually similar to other logged pages, PLS returns an SPT to the domain server. The domain owner saves this SPT and now returns the login pages to the client browser always with this SPT in a custom HTTP header. As described in Figure \ref{fig:architecture} the process begins when a user requests a webpage through the browser. The browser requests the domain server for the webpage. If the browser identifies that the page from a server is not a login page, it loads the page directly without page transparency checks. For login pages, the browser looks for a Signed Page Timestamp (SPT) header from the domain's server in an HTTP header and does a verification check to determine whether the page transparency is valid. If the SPT verification succeeds, the page loads. If it fails, the browser as a fall-back mechanism captures a screenshot and sends it, along with the domain name, to the PLS. The PLS computes an embedding of the screenshot, compares it with stored embeddings for the domain, and verifies the page's authenticity. The fallback mechanism helps prevent false-positive cases where small genuine modifications in authentic pages cause SPT failures. In case visual view of the logged page varies from stored one, an alert is generated to the user by the browser, otherwise, the page is safely loaded. Phishers cannot register their look-alike pages directly to PLS due to domain verification and look alike page detection mechanism. Also in case Phishers register a different page in PLS during registration and changes its view to a phishing page at a later stage are detected too via the fall back second check mechanism in browsers when SPT fails. Phishing is largely prevented and detected with the deployment of Page Transparency mechanisms proposed in this paper. 

The proposed architecture mandates that every login page that captures credential information from its users must be logged in one or more PLS. This allows legitimate login pages to obtain a SPT from the PLS, that any HTTP client can verify as proof that the page has been logged in its current visual form in one of the public log servers. The page's visual appearance is checked against other logged pages on the server before issuing an SPT. If it is visually similar to already logged pages from a different domain, the PLS will refuse to log the page to avoid phishers registering look-alike pages of targeted domains to the PLS.

The visual similarity between pages on the PLS is evaluated using the proposed Siamese Network Model (SNM). This model generates embeddings of each page and classifies them as visually similar, if the similarity score between the embeddings exceeds a predetermined threshold derived from experimental results. This ensures that two visually similar pages from different domains can never be logged by the PLS preventing phishers from registering Phishing pages on PLS. In cases where a phisher attempts to register a phishing page that is visually similar to a target page in order to receive an SPT, the model will detect the similarity during the registration process and prevent the page from being logged.

 Our login page detection algorithm that is proposed to be used by http clients such as browsers achieves a precision of 97.1\%, ensuring that nearly all login pages are correctly detected by the browser for SPT verification. Additionally, our architecture requires that the browser must always verify the SPT whenever a login page is rendered to the user. An SPT contains a cryptographic signature that is generated using a hash of the page's key textual content, signed by the PLS's private key to ensure authenticity, trust and integrity.

\noindent We also introduce an efficient browser-based re-verification algorithm to prevent potential false positives due to small genuine changes in the login pages. In the event of SPT verification failures, the browser can request the PLS to perform visual verification of the login page of the domain in real time using the SNM. The proposed architecture, inspired by the proven Certificate Transparency model, is the first end-to-end architecture specifically designed to address phishing attacks in an comprehensive architectural manner. 

\noindent The proposal work in two phases: the PLP and the Page Rendering Phase (PRP), described in sections \ref{sec:PLP} and \ref{sec:PRP}, respectively.
\subsection{\textbf{Page logging Phase (PLP)}}
\label{sec:PLP}
\begin{figure*}
\centering
\includegraphics[width=0.75\linewidth]{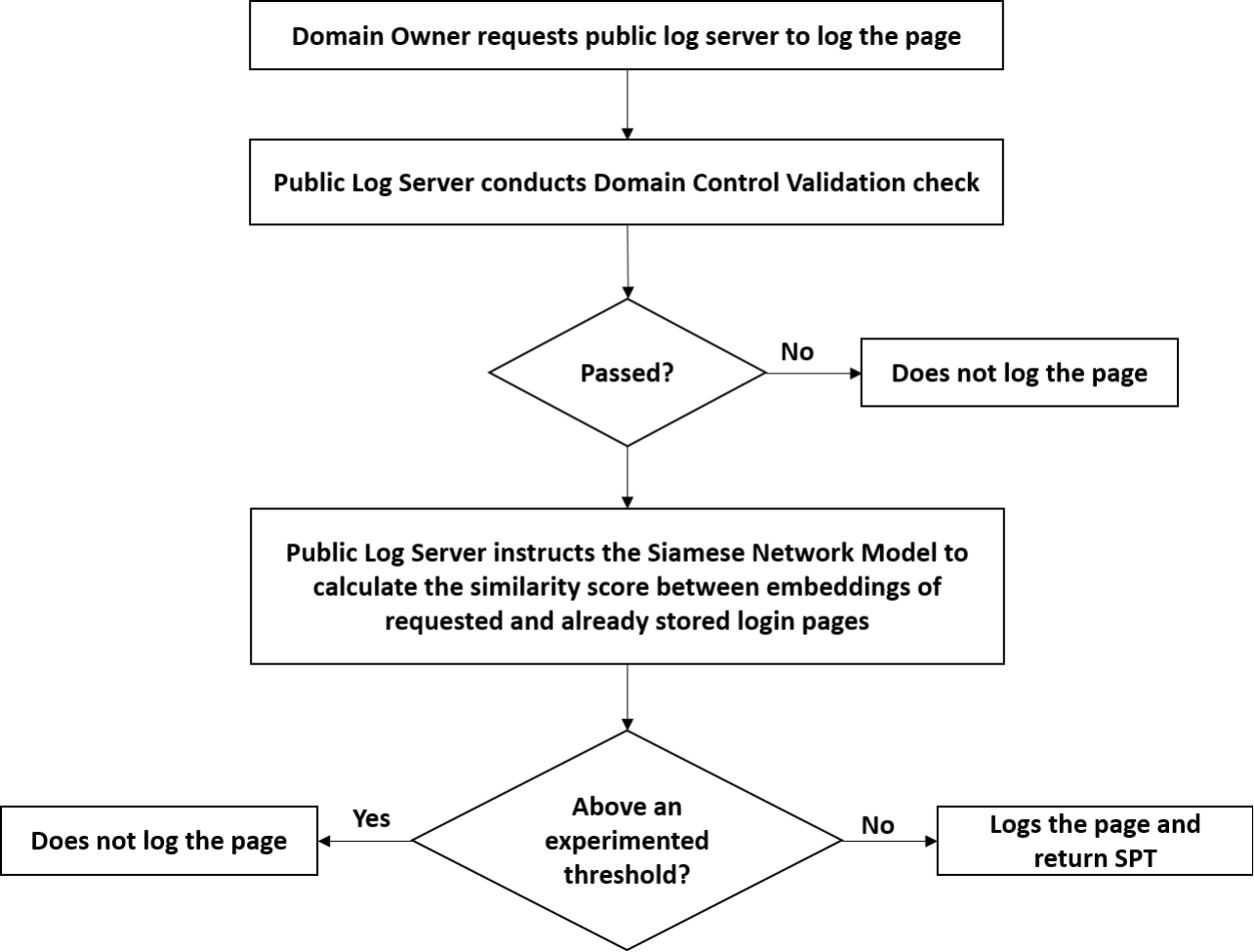}
\caption{\label{fig:phase-1}Page Logging Phase}
\end{figure*}

\vspace{0.5cm}
In the PLP, the domains register their login pages to PLS to receive an SPT. The PLS is equipped with an SNM trained on login pages of existing domains already registered in PLS. The PLS stores embeddings of previously registered login pages. Figure \ref{fig:phase-1} illustrates the step-by-step process of the PLP. To log a page, the domain owner initiates communication with the PLS by sending its certificate and a signature, signed by its private key. The PLS extracts the public key and domain name from the certificate and verifies the received signature using the extracted public key from the public key certificate. Successful domain verification and validation ensures that the login page belongs to the owner requesting the registration of the page in PLS for an SPT. The PLS then responds with a success message, along with its own certificate and a signature signed by its private key. The domain owner verifies the PLS certificate through certificate chain validation process used in X509 V3 certificate validations. Upon successful verification, it sends the page content and the URL to the PLS for page registration. The PLS then proceeds with its core responsibility.

The core responsibility of the PLS is to assess the visual similarity of the submitted login page with those already logged in the system. Leveraging the SNM trained for this purpose (explained in Section 3.1.1), the PLS evaluates the visual characteristics of the page. First, it computes the embeddings of the submitted page using the SNM. Then, it compares these embeddings to those of previously registered login pages (excluding pages from the same domain as the requester). If the page shows excessive similarity to other pages, exceeding a predetermined similarity threshold, it will not be logged. However, if the page is sufficiently distinct, it qualifies for logging, and the PLS proceeds to generate proof of its inclusion in the system.

To generate the proof, an SPT is generated using the HTML content and URL of the page. This timestamp generated is included in an HTTP custom header named SPT-Header, which the domain owner will include when responding the page to clients or browsers. This header serves as proof that the domain has got its page logged in the PLS. A detailed explanation of SPT generation is provided in section \ref{sec:SPT Generation}.

\subsubsection{\bfseries Siamese Network}

Motivated by \cite{2}, we embarked on the development of an SNM for visual similarity matching. SNM has gained prominence in various fields, particularly in computer vision tasks, due to their ability to assess similarity between pairs of inputs. In the context of Cybersecurity, detecting phishing websites remains a critical challenge due to the rapidly evolving nature of cyber threats. By utilizing visual similarity as a key indicator, our SNM aims to enhance the early detection of zero-day phishing websites. 

Following the methodology of \cite{2}, our dataset includes a comprehensive collection of 2,325 phishing, 115 legitimate, and 50 unseen legitimate login webpage screenshots, with a primary focus on Alexa's top websites that are the major victims of phishing attacks. This targeted approach enhances the model's ability to effectively detect and differentiate phishing websites based on visual similarity, especially in cases where user credentials are being targeted. \\

\begin{figure*}[ht]
\centering
\includegraphics[width=1\linewidth]{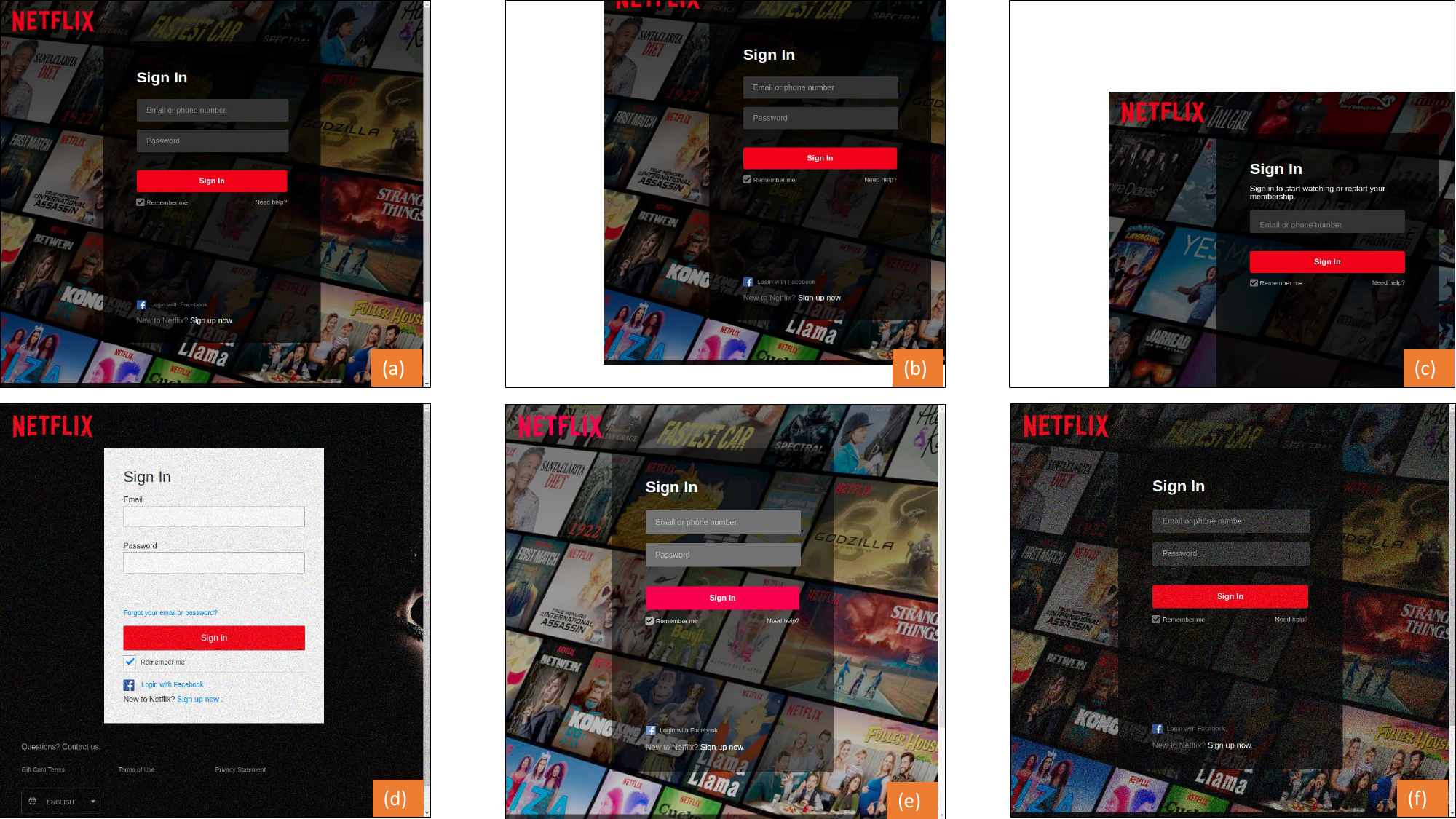}

\caption{In first row image \label{fig:amazon original} (a) representing original Sign In page of netflix, image (b) representing shift(-200,50), image (c) representing shift(-200,-200). In second row image (d) representing Gaussian Noise, image (e) representing Brightening and image (f) representing Darkening}
\end{figure*}

To train SNM, we created a dataset that includes screenshots of active verified phishing pages and their corresponding legitimate login page screenshots. Additionally, we incorporated a collection of unseen legitimate pages that do not match the legitimate pages corresponding to the phishing pages, for our testing and verification purposes. To improve the robustness of the model (so that small genuine variations does not generate false positives), we generated seven different variations of each phishing page. These variations include shifting the page towards its four corners, darkening and brightening the page, and adding Gaussian noise to the images. These seven variations are illustrated in Figure \ref{fig:amazon original}.
\begin{enumerate}
    \item 
We collected screenshots of phishing pages from PhishTank, a platform where phishing incidents are reported. Using an API key from PhishTank, we downloaded a JSON file containing the URLs of phishing pages and with Selenium WebDriver, we automated the process of capturing screenshots from these URLs. However, some of the phishing websites had already been blocked by the time they were reported on PhishTank, so screenshots for those websites could not be captured. In total, we successfully captured screenshots for approximately 1,500 pages initially, which belonged to various domains.

We then classified these pages according to their respective domains. Before doing so, we cleaned our dataset, which resulted in a reduced number of phishing pages. The screenshots were manually filtered in the following ways:

\begin{itemize}
    \item Some of the webpage screenshots captured appeared to be junk pages, displaying either black screen or pages that seemed disconnected from any recognizable domain.
\end{itemize}
\begin{itemize}
    \item Some of the webpage screenshots captured were not login pages. Since our solution is focused on login pages any other irrelevant pages were removed from consideration.
\end{itemize}

\item Collection of corresponding Legitimate website pages: 
We manually examined the filtered phishing pages to identify their corresponding domains. For each domain, we manually visited the legitimate website and searched for visually similar login pages. Once identified, we captured screenshots of these legitimate pages to include them in our dataset.
\begin{itemize}
\item Many of the domains for which phishing pages were captured no longer had functioning legitimate websites. This further reduced the overall size of our dataset.
\end{itemize}

\begin{itemize}
    \item Many of the legitimate login pages obtained had a different appearance compared to their corresponding phishing login pages. For instance, some platforms periodically update their login interfaces, leading to discrepancies between older versions and the latest iterations. In several cases, phishing pages closely resembled outdated legitimate login pages. To maintain the integrity of our dataset, we chose to exclude these outdated phishing page instances.

As a result, we created a dataset comprising 91 distinct domains, including 115 legitimate pages and 457 phishing pages.
\end{itemize}

\item For unseen legitimate pages we selected login pages from the top Alexa-ranked websites. We excluded the domains that were already used for training. In total, we created a dataset of 50 unseen legitimate pages from Alexa's top websites.

\item To enhance the model's robustness, we generated seven variations for each of the 457 phishing pages. These variations included shifting the pages to four corners, darkening, brightening, and adding Gaussian noise. After manually removing any variations unsuitable for training, we finalized a total of 2,325 phishing images.
\end{enumerate}
In this way, we created a dataset that includes 
\begin{itemize}
    \item 91 Distinct Domains 
    \item 2325 Phishing Pages (approximate average of 15 pages per domain)
    \item 115 Legitimate Pages (approximate average of 2 pages per domain)
    \item 50 Unseen Legitimate Pages (Legitimate pages that are not used for training)
\end{itemize}
\subsubsection{\textbf{Model Architecture}}
\noindent The architecture proposed in \cite{2} for similarity-based phishing detection employs a triplet network paradigm with shared convolutional networks. Each triplet comprises an anchor image, a positive image with the same identity as the anchor, and a negative image with a different identity. In our case, we have taken the legitimate login pages as anchor and their corresponding phishing pages as positive pages and the rest as negative pages. The objective is to learn a feature space where the distance between positive and anchor images' embeddings is smaller than the distance between anchor and negative images' embeddings. 

The training strategy involves two stages: initial training on all screenshots with random sampling, followed by fine-tuning with hard examples mined based on incorrect classifications. Triplet sampling is utilized, with triplets formed based on the L2 distance between embeddings. Minimum is the distance between the embeddings, more is the visual similarity between them. 

    In the testing phase, if the distance between a test screenshot and the trained embeddings of the legitimate websites is less than a predefined threshold, it signifies a match to the trained embeddings. For the specific purpose of phishing detection, any page classified as a match to the trained embeddings is identified as a phishing page, indicating an attempt to impersonate one of the trusted websites by having a high visual similarity. Therefore, a classification threshold is applied to the minimum distance to flag a page as a phishing page or a legitimate one with a genuine identity. This approach ensures that potential phishing pages are accurately identified based on their visual similarity to known legitimate websites. The structure of the SPT header is shown below:

\subsubsection{\textbf{SPT generation}}
\label{sec:SPT Generation}
\hfill\\
\begin{algorithm*}
\caption{SPT Generation}\label{alg:SPT}

\begin{algorithmic}[1]
\Function{Create\_SPT}{$\textit{url}, \textit{page}$}
    \State $\textit{hashed\_url} \gets \text{SHA256\_HASH}(\textit{url})$
    \State $\textit{hashed\_page} \gets \text{HASH\_HTML\_PAGE}(\textit{page})$
    \State $\textit{DER\_encoded\_public\_key} \gets \text{DER\_ENCODE}(\textit{PLS\_pem\_public\_key})$
    \State $\textit{struct\_data\_to\_sign} \gets \{\textit{version}, \textit{timestamp}, \textit{hashed\_url}, \textit{hashed\_page}\}$
    \State $\textit{signature} \gets \text{SIGN}(\textit{private\_key}, \textit{struct\_data\_to\_sign})$
    \State $\textit{SPT\_header\_data} \gets \text{concatenate\_bytes}(
    \text{bytes}(\textit{version}), \text{bytes}(\textit{timestamp}),$
    \State $\textit{stored\_logID}, \textit{signature})$
    \State $\textit{SPT\_header\_base64} \gets \text{Base64Encode}(\textit{SPT\_header\_data})$
    \State $\textit{SPT\_header} \gets \textit{SPT\_header\_base64}.\text{decode}('utf-8')$
    \State \textbf{return} $\textit{SPT\_header}$
\EndFunction
\end{algorithmic}
\end{algorithm*}
This section presents the algorithm for generating a Signed Page Timestamp. SPT serves as a cryptographic mechanism to provide verifiable timestamps for web pages.
\hfill\\
\noindent
Here, logID is a byte sequence derived from the SHA-256 hash of the DER-encoded public key of the PLS. These bytes representing logID are part of the SPT header data. This logID will be used as a unique identifier for each PLS. Logs containing logIDs and public keys of the PLS are publicly available. These logs can be used during SPT verification to obtain the public key of the PLS mapped to a logID in the logs. In the publicly available logs, logID will be presented as a Base64-encoded version of the hashed byte sequence. Figure \ref{fig:logID} shows how the representation of logID. \\
The version field indicates the version of the Page Transparency protocol. As of now, the PT protocol version is 1. The timestamp field in the SPT  represents the date and time at which the page was logged. 

The signature field is computed with parameters that include a hash of the page's textual content, a hash of the URL, version, and timestamp. These parameters are then signed by the PLS's private key to ensure authenticity, trust and integrity.

\hfill\\

\begin{tikzpicture}
    \draw[thick] (0,0) rectangle (2.5,-1) node[pos=.5] {\textbf{Version} \\ (1 byte)};
    \draw[thick] (2.5,0) rectangle (5.5,-1) node[pos=.5] {\textbf{Timestamp} \\ (4 bytes)};
    \draw[thick] (5.5,0) rectangle (8,-1) node[pos=.5] {\textbf{LogID} \\ (32 bytes)};
    
    \draw[thick] (0,-1) rectangle (8,-4.8);
    \node[anchor=north west, align=left] at (0.2,-1.2) {\textbf{Signature:}};
    \node[anchor=north west, align=left] at (0.5,-1.8) {
    \begin{minipage}{10cm} 
        \begin{itemize}
            \item Version (1 byte)
            \item Timestamp (4 bytes)
            \item URL Hash (64 bytes)
            \item Textual Hash (64 bytes)
        \end{itemize}
    \end{minipage}
    };
\end{tikzpicture}

\vspace{0.5cm}
\label{fig:sptheader}
\begin{figure}[ht]
\centering
\includegraphics[width=0.98\linewidth, height=9cm]{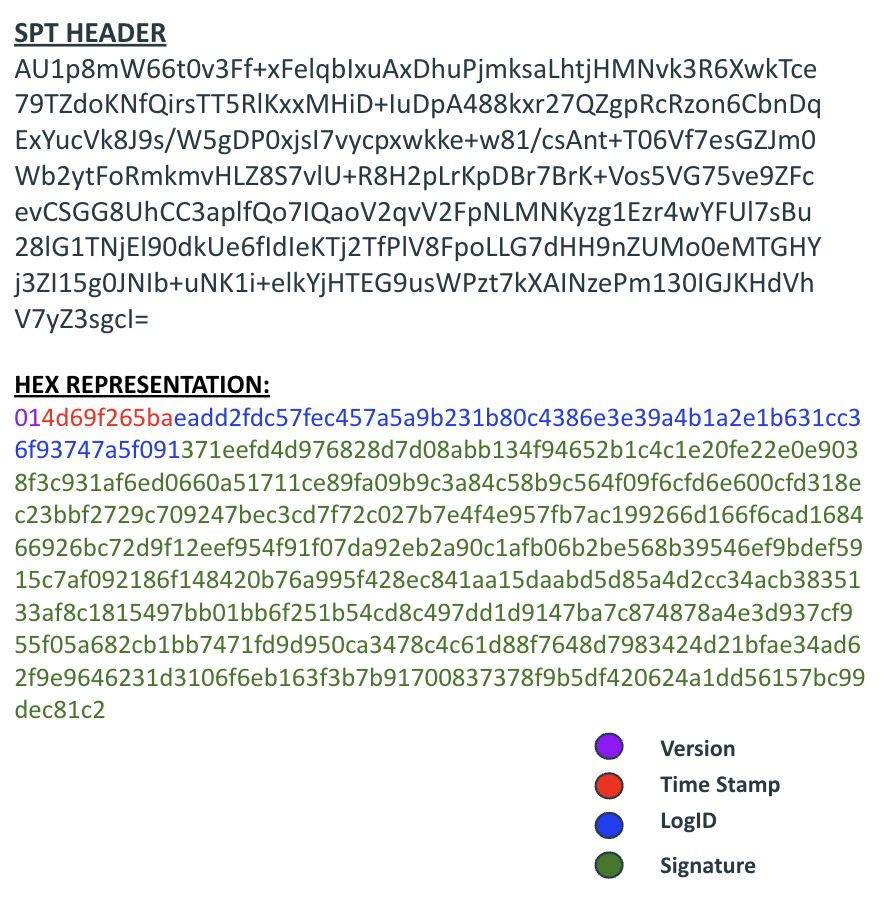}
\caption{\label{fig:spt_header2}SPT Header and its Hexadecimal representation }
\end{figure}
\hfill\\

\begin{figure*}[ht]
\centering
\includegraphics[width=0.9\linewidth]{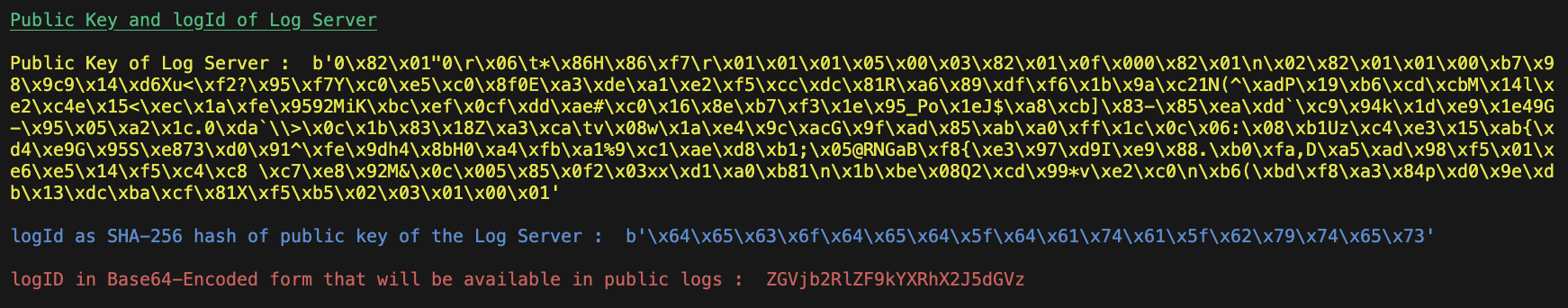}
\caption{\label{fig:logID}logID of PLS}
\end{figure*}
\newpage
The algorithm [\ref{alg:SPT}] describes the step-by-step procedure for generating a SPT. The CREATE\_SPT function takes the URL and HTML content of the page as inputs. Both the HTML content and URL are hashed using SHA256. A struct is then created using the version, timestamp, hashed URL, and hashed content.

Next, the struct is encrypted by the private key of the PLS, generating a signature. All the required bytes, including the version, timestamp, stored logID, and signature, are concatenated. Finally, these concatenated bytes are encoded to base64 and converted to UTF-8, forming the SPT Header. 
\hfill\\
%
\begin{table*}
    \centering
    \captionsetup{justification=centering}
    \caption{Abbreviation Table for Functions}
    \label{tab:Abbr}
    \begin{tabular}{|p{4.9cm}|p{4.25cm}|p{5.5cm}|}
    \hline
        \textbf{Function Name} & \textbf{Arguments} & \textbf{Definition} \\
    \hline
    
        SHA256\_HASH & \textbf{url}: url of the page & calculate sha256 hash of url \\
    \hline
        HASH\_HTML\_PAGE & \textbf{page}: HTML content of the page & calculates hash of a page\\ 
    \hline
        PACK\_DATA & \textbf{version}, \textbf{timestamp}, \textbf{hashed\_url}: SHA256\_hash(url), \textbf{hashed\_content}: SHA256\_Hash(html content) &  Creates a struct using all these parameters\\ 
    \hline
        DER\_ENCODE & \textbf{PLS\_pem\_public\_key} : Public key of PLS in PEM format &  Converts PLS public key from PEM format into DER Encoded PLS Public key\\ 
    \hline
        SIGN & \textbf{private key} : Private key of PLS, \textbf{struct data to sign} : struct(version, timestamp, hashed\_url, hashed\_page &  Signs struct data with private key of PLS to get signature\\ 
    \hline
        API\_CALL\_TO\_GET\_LOGS & \- &  Fetches all stored logs from PLS\\ 
    \hline
        SIGNATURE\_VERIFICATION & \textbf{spt\_header}, \textbf{url}, \textbf{content} & This algorithm fetches log\_id from spt\_header, check if any log exists with decoded form of this log id, and if it exists, it verified the signature obtained from spt\_header using public key of the PLS. \\
    \hline
        IS\_LOGIN\_PAGE & \textbf{html\_content} & This algorithm looks into important aspects of html content and return whether the page is a login page or not  \\ 
    \hline
        CHECK\_PHISHING\_OR\_NOT- \_USING\_MODEL
        & \textbf{screenshot} : Screenshot of the page being loaded by the user & Calculates embedding of the page and compares with embeddings of existing stored pages to check if current page is phishing page of any existing legitimate page\\ 
     \hline
    \end{tabular}
\end{table*}
\begin{table}[ht]
    \centering
    \caption{Abbreviation Table for Variables}
    \label{tab:my_label}
   \begin{tabular}{|p{3.1cm}|p{4.2cm}|}
    \hline
        \textbf{Variable Name} & \textbf{Definition} \\
    \hline
        loginThreshold & This is the value, above which if a page has a login score, then that page is considered as a login page. We have set it as 76. \\
    \hline
        siameseThreshold & This is the value, below which if we have the difference between embeddings, then the pages are considered to be similiar and one can be said to be phishing page of another.\\
    \hline
        shouldContinueLoading & This is a boolean value, that decides weather the page should be continued to be loaded, on doing all the checks to detect it as phishing page\\
    \hline
    \end{tabular} 
\end{table}
\subsection{\textbf{Page Rendering Phase}}
\label{sec:PRP}
\begin{figure*}
\centering
\includegraphics[width=1\textwidth, height=0.55\textheight]{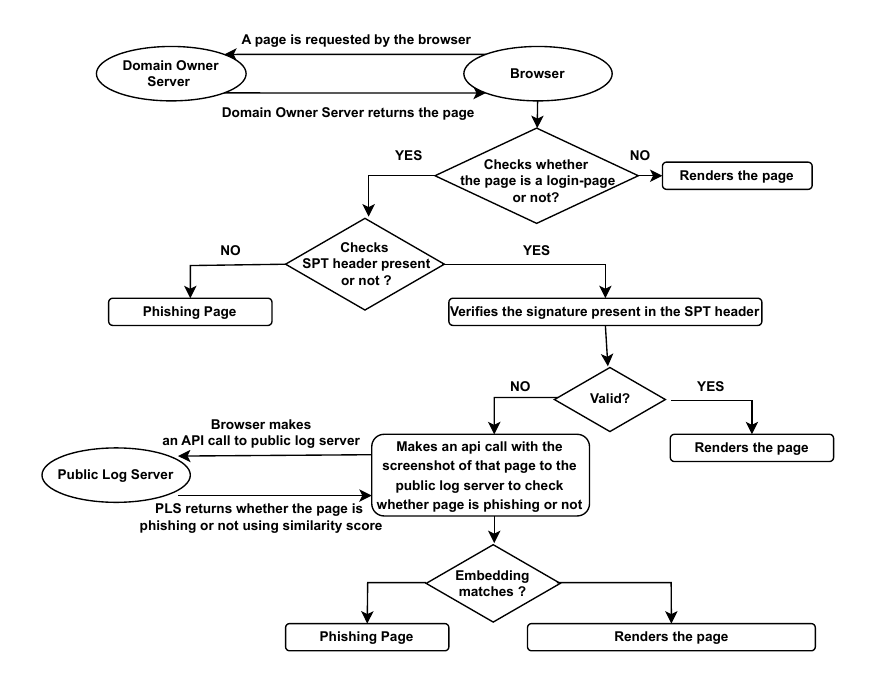}
\caption{\label{fig:phase-2}Page Rendering Phase}
\end{figure*}
In \textbf{Page Rendering Phase}, there exist three entities, a domain owner server (DS) that serves a page requested by the browser, a browser that performs signature (present in SPT) validation before rendering any login page, and a PLS that is used to verify the authenticity of the page served by the domain owner in case browser is unable to verify a login page using the SPT. When a user requests a login page, the domain owner returns the page along with an SPT in a custom HTTP header. The browser retrieves the SPT header and verifies the authenticity of the page using our proposed SPT verification algorithm that takes the current form of the web page as input along with the information present in SPT to verify the signature. A login page without an SPT header is not rendered by the browser. In the case when the SPT verification fails, the browser will send the visual screenshot of the page to the PLS to classify the page as phishing or legitimate based on whether the same visual page is registered in the PLS against the domain. If the page embedding stored in the PLS against the domain does not match with the visual screenshot embeddings, the page will not be rendered. This avoids phishing pages targeting a specific domain to be completely unavailable for interaction with a victim on the browsers.

To get a page identified as a login page at the browsers we have developed a mechanism to detect login/signin/signup pages which is discussed in leading paragraph. If it is not a login page, the browser proceeds to render the page normally. However, if it is indeed a login page, the browser then looks for the presence of the SPT header. In the absence of the SPT Header, the page is flagged as a potential phishing attempt and is not rendered. However, if the SPT header is present, the browser proceeds to verify its signature.

During signature verification, if the signature is deemed valid, the page is confirmed as legitimate and is rendered for the user. The detailed information and algorithm for signature verification are discussed in section \ref{sec:SV}. However, if the signature validation fails, we have implemented an additional safeguard mechanism to prevent potential Denial of Service. In this scenario, the browser captures a screenshot of the suspicious page and makes an API call to a designated PLS.

The PLS analyzes the received screenshot to determine whether the page is indeed phishing or not. If the PLS confirms the page as phishing, it instructs the browser not to render the page, thus protecting the user from potential harm. Conversely, if the PLS does not identify the page as phishing, it is deemed legitimate and is rendered for the user's access.

A flowchart illustrating the Page Rendering Phase has been included for visual clarity. This flowchart outlines each step of the process, including the verification of the presence of the SPT header, signature validation, and the secondary check for potential phishing attempts.

\begin{algorithm*}
\caption{Login Page Detection Algorithm}
\label{algo:LPDA}
\begin{algorithmic}[1]
     \Function{is\_Login\_Page}{html\_content}
            \State $score \gets 0$
            
            \State for each keyword in \textit{loginKeywords}
                \State \hspace{0.5cm} if keyword in HTML content
                    \State \hspace{1cm} $score \mathrel{+}= 10$
                    
            \State for each keyword in ["signin", "signup", "login", "log-in", "sign-in", "sign-up"]
                \State \hspace{0.5cm} if keyword in URL
                    \State \hspace{1cm}  $score \mathrel{+}= 30$
                    \State \hspace{1cm} break
            \State if Submit button is present
                \State \hspace{0.5cm} $score \mathrel{+}= 15$
            \State if input fields are present
                \State \hspace{0.5cm} for each input field
                     \State \hspace{2em} \textbf{if} input field contains 
\State \hspace{3em} [name="username", name="userid", name="email", 
\State \hspace{3em} type="email", type="password", 
\State \hspace{3em} placeholder="username", placeholder="email", placeholder="password"]

                        \State \hspace{1.3cm} $score \mathrel{+}= 60$
            \State if $score > \text{loginThreshold}$
                \State \hspace{0.5cm} return true
            \State else
                \State \hspace{0.5cm} return false
    \EndFunction
\end{algorithmic}
\end{algorithm*}

As Phishers commonly employ deceptive tactics to obtain users' login credentials via look-alike target domain login/signin forms, our scheme only come into action if a page visited by a user is a login page. Hence, we validate the presence of a legitimate login form on a webpage before moving on to subsequent steps of our phishing detection methodology. Should no login forms be detected, further detection procedures are deemed unnecessary, as users lack a means to input their confidential information.

Our research refine the conventional login page detection algorithm through a novel approach utilizing weighted analysis of login page identifiers. Drawing inspiration from technical reports such as Invicti's "Sitemap Analysis" \cite{37}, we developed a methodology that adjusts weights of login identifiers based on empirical observations from our dataset and iterative experimentation. The identifiers included login keyword in HTML content, login keyword in window location, submit button, inputTags. We assigned different weights to each of the identifiers. We check for different identifiers found on the page and summed up their weights, then classified it as a login page if the total weight crosses a threshold, otherwise classified it as a non login page. We have kept the threshold at 75 based on the experimentation we did with 105 websites. This check can be configured, by increasing or decreasing the weight of variables, such as password input and/or adding new keywords in the future.

In our approach, we assigned higher weight to key indicators such as input tags with name/type/placeholder attributes having values like email/username/password/userId , prioritizing their significance in capturing essential login credentials. Following it, URLs containing login-related keywords received moderate weighting. The presence of a submit button with type="submit" attribute further contributed to the overall weight of 15. Then we had login keyword check for html content, where presence of each keyword contributed a weight of 10. Initially we had assigned larger weight to this, but based on an observation we reduced the weight of each keyword to 10. The observation was that in some of the news websites, they included a lot of content, and it was more likely to find the login keywords on that site. So we reduced the contribution of each keyword in declaring the page as a login page. Also, for the same reason, we introduced an upper limit on the cumulative weight attributed to keywords.
Here is more description about the identifiers we had used in our algorithm.
\begin{itemize}
    \item Weight of login keyword in HTML element: We define a list of 39 login keywords 
    
    login keywords = [
        "username", "password", "login", "signin", "sign-in", "log in", "log-in", "authenticate", "credentials", 
        "account", "identity", "user", "email", "e-mail", "passcode", "customer number", 
        "pin", "secret code", "authentication code", "security code", "passphrase", "account number", 
        "membership number", "social security number", "authorization code", "login code", 
        "secure login", "unique identifier", "login id", "login name", "login details", "login information", 
        "login credentials", "login data", "login token", "login key", "userid", "forgot", "Log in", "Login", "Email", "Username", 
        "Sign in","signed in", "Phone", "phone"
    ]
    
    These keywords are most
commonly found on Login/SignUp pages. We check how many of these keywords are present in HTML content of the page.
Weight Assigned: 10/each word

\item Weight of login keyword in window location: We check if any of the keywords like login, log-in, Log in
signin, sign in, sign-in, Sign up is present inside the URL of the page.
Weight Assigned: 30

\item Weight of submit button: We check if any submit button is present on the page or not.
Weight Assigned: 15

\item Weight of inputTags: We check if any of the input tags are present in form tags that have type=
username /email / password/ userid or placeholder includes words like email/password/ userid/ username.

Weight Assigned: 60
\end{itemize}
The pseudo code for the login page detection algorithm can be found in Algorithm \ref{algo:LPDA}.

\subsubsection{Signature Verification}
\label{sec:SV}
\hfill\\
If the page is a login page, then from the HTTP headers the SPT header is extracted. Version, timestamp, logID, and signature are obtained from the SPT header. The logID is checked in the list of publicly available logs. If the logID does not match any of the logIDs, it signifies that the page is not logged in any of the trusted PLS. In that case, the page is classified as a phishing page. For a valid logID, the public key of the corresponding PLS is extracted. This public key will be used to verify the digital signature. The version, timestamp, hashed URL, and hashed HTML content is packed in a structure and the SHA-256 Hash is computed over the structured data. Finally, by using the obtained public key of the PLS, the computed hash and the digital signature present in the SPT Header, the SPT is verified. If the signature is verified, it confirms that the signature is generated by an authentic PLS and the page is logged in that server in its current form. If the signature verification fails, it means that after the login page corresponding to that domain was logged into the PLS, the page has changed, and that the signature created using its current HTML content does not match the signature it was logged with. This signature verification is explained algorithmically in Algorithm \ref{algo:SVA}.
\begin{algorithm*}
    \caption{Signature Verification Algorithm}
    \label{algo:SVA}
    \SetAlgoLined
    
    \SetKwFunction{FHashText}{\string hash\_text}
    \SetKwFunction{FBaseDecode}{\string base64.decode}
    \SetKwFunction{FBaseEncode}{\string base64.encode}
    \SetKwFunction{FPackData}{\string pack\_data}
    \SetKwFunction{FSHAhash}{\string SHA256\_hash}
    \SetKwFunction{FDerEncode}{\string der\_encode}
    \SetKwFunction{FAPICall}{\string api\_call\_to\_get\_logs}
    
    \SetKwProg{Fn}{function}{}{}
    \Fn{SIGNATURE\_VERIFICATION(spt\_header, url, content)}{
        hashed\_url $\gets$ \FHashText{url}\;
        hashed\_content $\gets$ \FHashText{content}\;
    
        decoded\_spt\_header $\gets$ \FBaseDecode{spt\_header}\;
        version $\gets$ decoded\_spt\_header[:1]\;
        timestamp $\gets$ decoded\_spt\_header[1:5]\;
        log\_id\_bytes $\gets$ decoded\_spt\_header[5:37]\;
        signature $\gets$ decoded\_spt\_header[37:]\;
    
        packed\_data $\gets$ \FPackData{version, timestamp, hashed\_url, hashed\_content}\;
        hashed\_data $\gets$ {\FSHAhash{packed\_data}}\;
    
        log\_id $\gets$ \FBaseEncode{log\_id\_bytes}\;
        logID\_found $\gets$ \textbf{False}\;
        index $\gets$ 0\;
        log $\gets$ \FAPICall{}\;
    
        \For{$i \gets 0$ \KwTo $\texttt{len}(log) - 1$}{
            \If{log\_id == log[$i$]["log\_id"]}{
                logID\_found $\gets$ \textbf{True}\;
                index $\gets$ $i$\;
                \textbf{break}\;
            }
        }
    
        \If{\textbf{not} logID\_found}{
            \textbf{return} \textbf{False}\;
        }
    
        public\_key $\gets$ log[index]["pub\_key"]\;
        der\_encoded\_public\_key $\gets$ \FDerEncode{public\_key}\;
    
        \If{der\_encoded\_public\_key.verify(signature, hashed\_data)}{
            \textbf{return} \textbf{True}\;
        }
        \Else{
            \textbf{return} \textbf{False}\;
        }
    }  
\end{algorithm*}

\begin{figure*}[ht]
\centering
\includegraphics[width=1\linewidth]{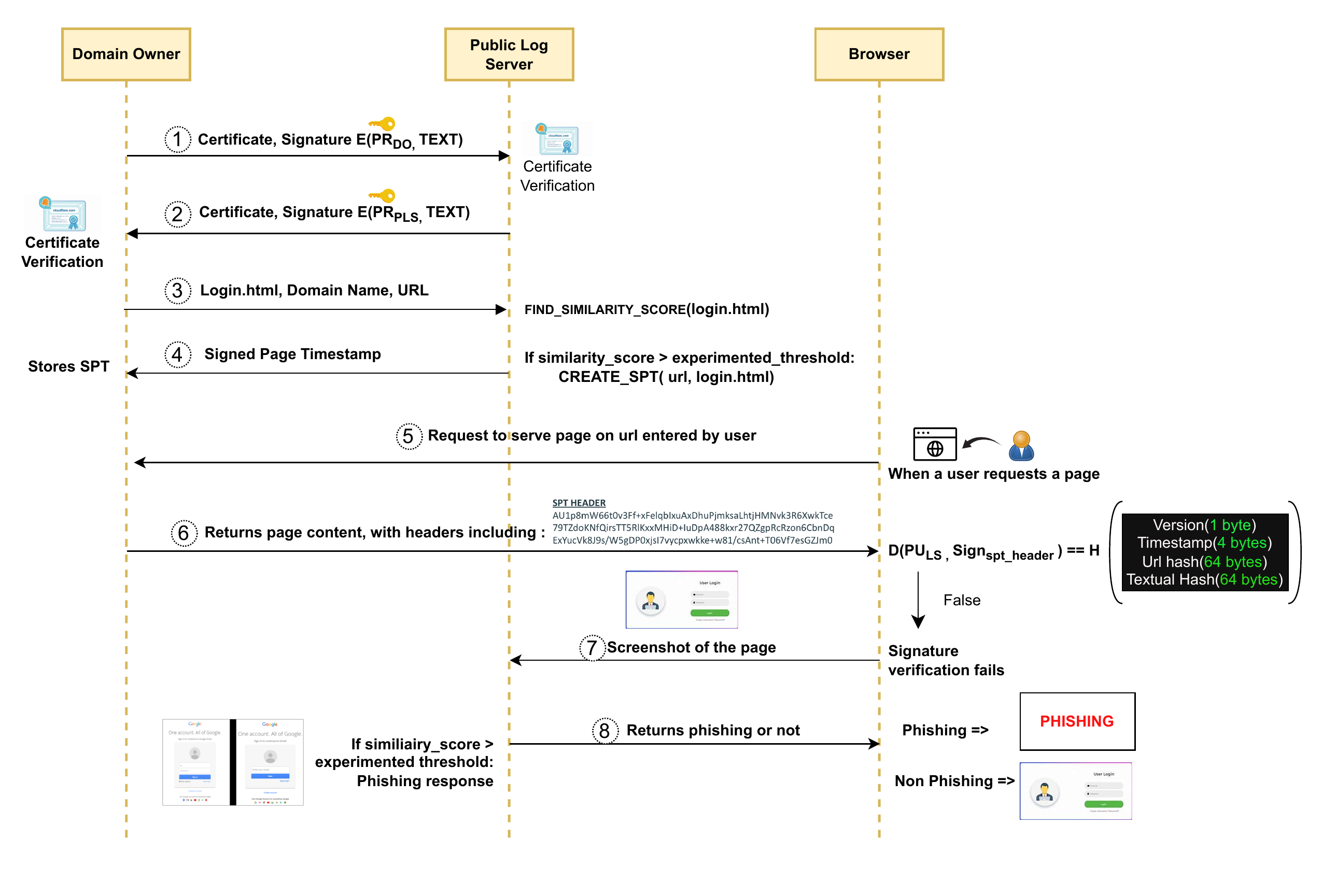}
\caption{\label{fig:Flow_Diagram3}Comprehensive communication among all entities}
\end{figure*}

\hfill\\
We have incorporated the whole flow of both phases in Figure \ref{fig:Flow_Diagram3}.
\section{Experimental Implementation and Deployment of the Architecture}
Since we couldn't modify core functionality of a commercial browser or would have been time consuming, we developed a browser extension named "AntiPhish" to mimic the proposed additional functionalities. This extension developed alters how the browser handles the page request, allowing us to incorporate our architecture into it, which can be later adopted by any browser in their core functionality. We created this extension for Google Chrome v 124.0.6367.119, which works with Manifest v3.
The proposed system was implemented on an Ubuntu 22.04.4 LTS OS running on a Z2 Tower G5 Workstation with an Intel Core i5-10500 CPU @ 3.10GHz, 32 GB of RAM, and 2 TB of disk capacity. For the Flask servers and Python applications, the following versions were utilized: Flask (Version 3.0.0), Python (Version 3.10), cryptography (Version 41.0.5), numpy (Version 1.19.5), tensorflow\_gpu (Version 1.14.0), matplotlib (Version 3.3.4), etc. \\

 Our deployment architecture comprises three entities: the PLS, the Browser, and the Domain Owner. Figure \ref{fig:Flow_Diagram3} represents the comprehensive communication among these entities.
\begin{enumerate}
\item The domain owner initiates communication with the PLS by sending its certificate and signature signed by its own private key. The PLS extracts the public key and domain name from the certificate and then verifies the received signature using the extracted public key. Successful verification ensures that the domain belongs to the requester.
 \item The PLS respond with its certificate and signature signed by its private key. The domain owner verifies the certificate and the signature. 
\item After verification is done via challenge-reponse on both ends, domain owner send HTML file of its login page alongwith domain name and URL, to PLS. PLS is already equipped with the functionality of visual similarity matching of the requested login page with stored embeddings of the other domains using SNM. 
\item If the current page exhibits excessive resemblance to existing logged pages, surpassing an experimented threshold of the similarity score, the page will not be logged in the PLS. If the page demonstrates sufficient distinctiveness, it qualifies for logging and the PLS proceeds to generate proof of its inclusion in the system. For generating such proof, a signed page timestamp is created by the PLS by hashing HTML content, URL and other specifics as per SPT generation algorithm. This domain owner stores the SPT returned by PLS.
\item Whenever a user visits the login page of the domain he/she enters a URL in browser or make a web request to the server.
\item The domain owner returns page content along with HTTP headers The SPT header is extracted from the headers. Version, timestamp, logID, and signature obtained from the SPT header are used to verify the signature.
\item If the signature verification fails, then browser sends the screenshot of the page along with the url to PLS as a second check. 
\item At PLS, embedding of the received screenshot is compared against the stored embeddings of the logged pages.
If the similarity score with any logged page is greater than the experimentally obtained threshold, domain name of the logged page is matched against the url sent with the screenshot. If the domain does not match, it is inferred that the page served to the user attempts to impersonate one of the trusted websites. It is declared as phishing and the page will be stopped from loading on the user browser. If the domain name matches, the screenshot is considered to be a light variation of a trusted logged website and is declared safe.
\end{enumerate}
A more detailed implementation of the Antiphish extension can be found in the Appendix \ref{app:A}.

\begin{algorithm*}[ht]
\caption{Working of Browser Extension Used for Prototype Deployment}
\label{algo: Extesion Working}
\begin{algorithmic}[1]
    \Function{Event Listener for chrome.webRequest.onHeadersReceived}{}
        \State \hspace{1em} headers = response headers
    \EndFunction
    
    \Function{Event Listener for chrome.webNavigation.onDOMContentLoaded}{}
        \State \hspace{1em} shouldContinueLoading = true
        \State \hspace{1em} isLogin  = Function(IS\_LOGIN\_PAGE(html content))
        
        \State \hspace{1em} if (isLogin == True)
            \State \hspace{2em} sigVerified = SIGNATURE\_VERIFICATION(SPT\_HEADER, url, html content)
            
            \State \hspace{2em} if (sigVerified == False)
                \State \hspace{3em} SS = Captured Screenshot of Page
                \State \hspace{3em} isPhishing = CHECK\_PHISHING\_OR\_NOT\_USING\_MODEL(screenshot)
                
                \State \hspace{3em} if isPhishing == True
                    \State\hspace{4em} shouldContinueLoading = False
         
        
        \State \hspace{3em} if (shouldContinueLoading == false)
            \State \hspace{4em} Stop loading the page and display a phishing warning
    \EndFunction
\end{algorithmic}
\end{algorithm*}
\begin{algorithm*}
\caption{Server Deploying SNM Model}
\label{algo: Model Server}

\SetAlgoLined

\SetKwFunction{FMain}{CHECK\_PHISHING\_OR\_NOT\_USING\_MODEL}
\SetKwProg{Fn}{function}{:}{}
\Fn{\FMain{$screenshot$}}{
    new\_embedding = calculate\_emb($screenshot$)\;
    
    \For{exist\_embedding \textbf{in} database}{
        \If{diff($exist\_embedding$, $new\_embedding$) $<$ siameseThreshold}{
            \Return true \tcp*{is phishing}
        }
    }
    \Return false \tcp*{not a phishing page}
}
\end{algorithm*}
\section{Threat Model}
\begin{table*}[ht]
\scriptsize
    \centering
    \caption{Threat Model} \label{tab:threat_model}
    \begin{tabular}{|p{2cm}|p{6cm}|p{7.5cm}|}
    \hline
        \textbf{Attack Name} & \textbf{Threat} & \textbf{Justification} \\
    \hline
        Cross-Domain Request Attack & An attacker may attempt to send a request on behalf of another domain to log a page in the public server. & This attack is mitigated by involving certificate exchange. Each domain is required to provide a valid certificate and a challenge signed by the certificates private key, ensuring the authenticity of the request. Only domains with valid certificates can log pages, preventing unauthorized requests. \\
    \hline
        Header Spoofing Attack (SPT) & An attacker could try to add a spoofed SPT as a header. & The SPT verification mechanism ensures that only valid SPTs are accepted. Publicly available logs of public keys associated with a PLS allow browsers to verify that the signature in an SPT was created using a trusted PLS. \\
    \hline
        Visual Similarity Evasion & An attacker might create variations of target login web pages and request PLS to log them to obtain an SPT. &  Our architecture employs a SNM trained to recognize different variations of web pages as similar. This model ensures variations are detected and appropriately handled, preventing exploitation through subtle visual changes in a web page. The model can correctly detect a smart variation of the page by an attacker and deny it for logging over PLS. \\ 
    \hline
        SPT Verification Failure & If a legitimate website's SPT verification fails, it could lead to a denial of service for that website. &  The architecture has a backup process for handling SPT verification failures. If an SPT verification fails, the request is forwarded to the PLS for visual similarity verification using SNM. This ensures that legitimate websites are not denied service due to temporary verification issues, maintaining continuous service availability. \\ 
    \hline
         Phishing Page Rendering Bypass &The page will be scrutinized by the browser only if it is a login page otherwise it will be rendered without any check. An attacker can render its login phishing pages by bypassing login-page detection mechanisms. &  This attack is mitigated using our proposed advanced login page detection algorithm in which proper weights are given to certain keywords according to their importance. \\ 
    \hline
         Denial of Service (DoS) & If the SPT verification fails, then the request gets forwarded to the PLS for model check. Overwhelming the browser with malicious pages with spoofed SPT, can disrupt the service.  &  PLS redundancy and necessary rate limiting will protect such occurences. \\
    \hline
         PLS Compromise & If the PLS is compromised, logged pages could be tampered with or erased. &  NA \\ 

    \hline
    \end{tabular}
    
\end{table*}
To ensure the security and integrity of the proposed architecture for logging web pages on a public server, it is important to consider and address potential threats and vulnerabilities. We will analyze the threat aspect of each entity in our architecture. The primary entities involved in our architecture include: Browser, PLS, Domain Owner.
\begin{itemize}
  
\item \textbf{Browser:}
The threats on browsers majorly include by phishing page rendering bypass, header spoofing attacks, and denial of service (DoS) attacks. Bypass of a phishing detection check on a page is mitigated through an advanced login page detection algorithm. As the phishing page will demand a credential and will be detected by the login detection algorithm and hence cannot be bypassed against further checks. Header spoofing attacks are countered by verification of SPT headers against public logs which are also digitally signed. A spoofed SPT header will get easily detected. To prevent DoS attacks on browsers, rate limiting, CAPTCHAs, load balancing, and redundancy are implemented to manage high traffic and ensure continuous service.

\item \textbf{PLS:}
The PLS faces threats such as cross-domain page log request attacks, visual similarity evasion, and potential compromise. Cross-domain request attacks are mitigated by requiring valid certificates for domain authenticity before logging a specific page against a domain in the PLS. Visual similarity evasion is handled by an SNM trained to detect subtle variations in web pages with a very good accuract. To protect against server compromise, cryptographic signatures, redundancy with multiple PLS, and regular audits can be utilized during deployments.

\item \textbf{Domain Owner:}
Domain owners are threatened by SPT verification failure and visual similarity evasion. SPT verification failure, which could lead to service denial, is mitigated by forwarding failed verifications to the PLS for backup verification using the SNM. Visual similarity evasion, where attackers use slightly altered phishing pages, is addressed by the SNM to ensure variations are detected and handled properly and the phishing page that are visually similar to domain owner's page are not logged over the PLS.

The detailed attack and mitigation strategy is discussed in Table \ref{tab:threat_model}. \\
\end{itemize}
\section{Experimental Results and Analysis}
\subsection{Login Page Detection Algorithm}
We tested our proposed login page detection algorithm over 103 unique webpages that had banking sites, news websites, video streaming, web hosting websites, learning platforms, etc. Some of the major domains we tested our algorithm on are given below. 

[microsoft, google, mailchimp, shopify, wordpress, twitter, shift4shop,
facebook, dropbox, trello, salesforce, ebay, paypal, etsy, godaddy, quora,
icicibank, udemy, stackoverflow, cricbuzz, jiosaavn, github, tribun-
news, twitch, netflix, instagram, imdb, imgur, booking.com, bbb.com,
cnn.com, onlinesbi.sbi, roblox, aws.amazon.com, chase.com, mediafire, mytimes.com, stackexchange,  indeed, researchgate.net, wetransfer.com,
w3schools.com, theguardian.com,  ndtv, walmart.com, dailymotion.com, steam-
community.com, timesofindia, saral.iitjammu, indianexpress,  
hindustantimes, bookmyshow, zomato, swiggy, bing.com, tribune-
news.com, blogger.com, msn.com, zoom.us, soundcloud.com]

We had 67 Login/SignIn/SignUp pages and 36 non-login pages. These were the various metrics we obtained.
TP:67, TN:34, FP:2, FN:0. Precision of our login page detection algorithm comes to be equal to 0.971 and accuracy comes as 0.981.

\subsection {Siamese Network Model}
For training our model, we used 115 legitimate images and 2325 phishing images. We allocated 30\% of the phishing login page images for testing, resulting in 697 phishing images for testing and the remaining 1628 for training. Combining the training set of phishing images with the legitimate images, we had a total of 1743 training images.  Our testing set consists of 747 images in total, with 697 images expected to be classified as phishing and 50 additional legitimate login page images as non-phishing.

\begin{figure}
    \centering
    \includegraphics[width=1\linewidth]{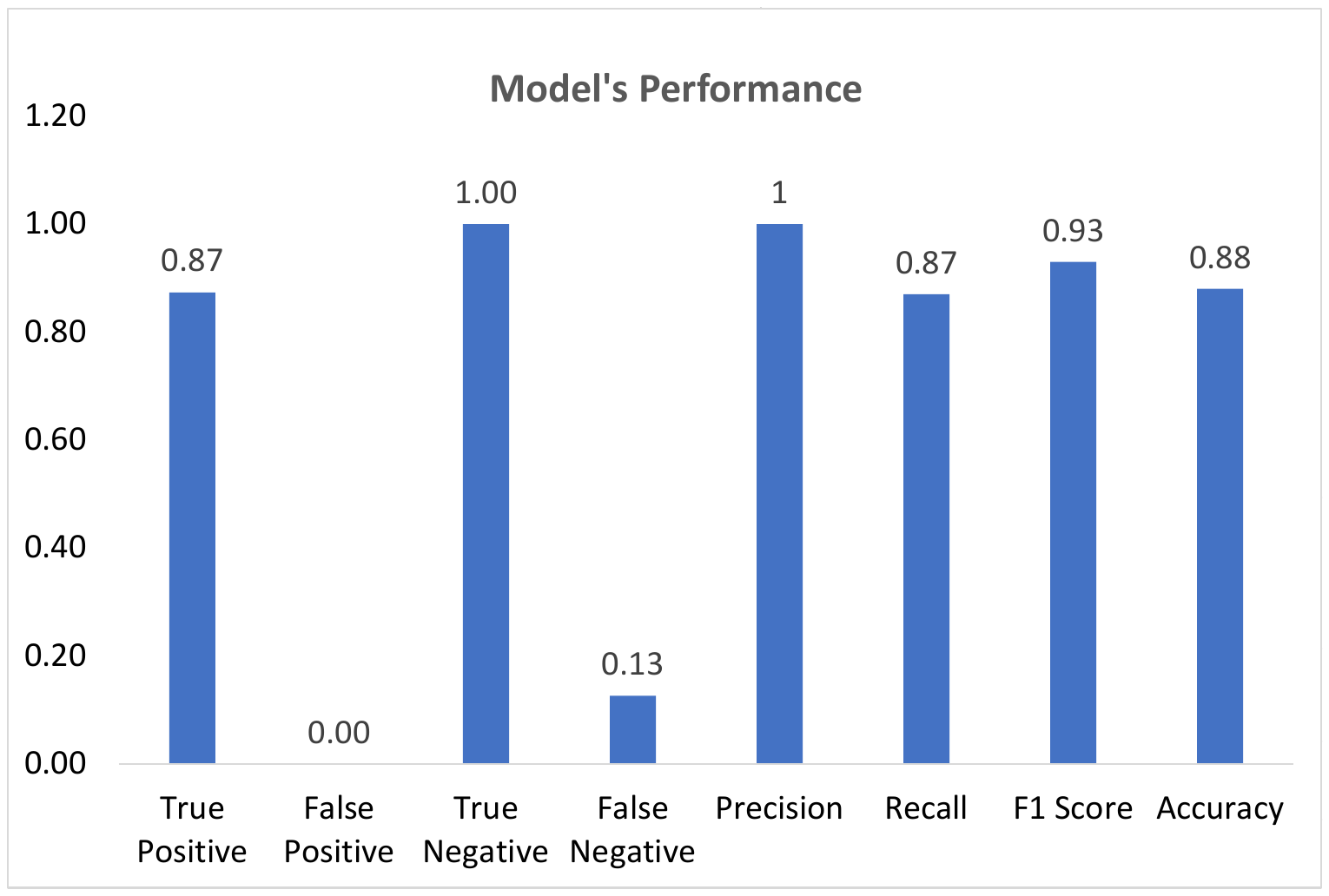}
    \caption{Performance parameters out of 697 phishing and 50 unseen legitimate images}
    \label{fig: phishing sites}
\end{figure}

In our analysis, as described in Figure \ref{fig: phishing sites}, phishing images are considered positive, and unseen legitimate images are considered negative. Using our trained model over the training dataset described earlier and with an experimented threshold, out of the 697 phishing images, 609 are correctly identified as phishing, resulting in 609 true positives (TP). The remaining 88 phishing images are incorrectly identified as non-phishing, resulting in 88 false negatives (FN). For the unseen legitimate images, all 50 are correctly identified as legitimate since the distance between these unseen legitimate image embeddings and the trained embeddings exceeds the experimentally determined threshold. This resulted in FP of 0. We finally received a Precision of 1, a Recall of 0.873, and an F1 Score of 0.932. 


\begingroup
\subsection{Performance of Model on local region-specific pages}
To thoroughly evaluate the robustness and generalizability of our approach, we undertook significant enhancements to both our dataset and adversarial sample generation methods. Recognizing the importance of representing real-world scenarios, we aimed to cover a diverse range of brands and attack techniques. The following section details the expansion of our dataset to include both prominent global brands and local entities, as well as the innovative strategies employed to create challenging adversarial examples for comprehensive performance assessment. \\
 We expanded the dataset to include a range of local and global brands, covering both popular and less well-known names such as Facebook, X, Instagram, GitHub, Paytm, Netflix, LinkedIn, Google, Nykaa, AWS, Azure, Flipkart, Amazon, Zepto, Blinkit, Google Play, Disney+ Hotstar, Swiggy, Zomato, PhonePe, Jio, Prime Video, YouTube, Snapchat, IRCTC, RedBus, Goibibo, MakeMyTrip, BookMyShow, Meesho, Rapido, Temu, Tumblr, VK, Walmart, WeChat, Wells Fargo, Yahoo, Yandex, OnlineSBI, PayPal, Reddit, Shopify, Snapdeal, SoundCloud, Spotify, Target, HSBC, Hulu, ICICI Bank, Myntra, Disney+, Drugs.com, DuckDuckGo, Etsy, Goldman Sachs, HDFC Bank, Axis Bank, Baidu, Best Buy, Bing, Chase, Citi, Crunchyroll, Dailymotion, American Express, and others. These brands represent a diverse range of industries, including educational platforms, entertainment, news, social media, fashion, banking, e-commerce, shopping, and travel websites. Our dataset is focused solely on login pages, as these are more susceptible to phishing attacks.\\
To create adversarial samples, we adopted a new approach in addition to the previous adversarial sample generation methods discussed in this paper. This time, we designed a generic login form layout commonly used by many websites. We first select the logo and primary color of the original website, and then apply that color to our generic template to generate adversarial samples. The login form can be placed in nine different positions on the webpage—top left, top center, top right, middle left, middle center, middle right, bottom left, bottom center, and bottom right—to further diversify the samples. These layouts generally do not arouse suspicion easily.
For additional adversarial samples, we followed another approach: replacing the current logo on the website with one of its previous official logos. Since these older logos were familiar to users in the past, they can still effectively deceive people. In this case, we replicate the current webpage template but substitute the current logo with a selection of previous logos.
\subsection{Mobile Specific Pages}
In addition to desktop-oriented login pages, we extended our analysis to mobile-specific variants. For this purpose, we constructed a mobile-specific dataset, comprising mobile login pages collected from the responsive versions of major websites. Each page was rendered and captured using mobile viewport dimensions, emulating common mobile screen sizes and aspect ratios.  \\
The dataset includes websites from a diverse range of sectors such as social media, e-commerce, cloud storage, banking, streaming services, and developer platforms. The following websites were used:
Bbb, Bing, Dailymotion, Hindustantimes, Imdb, Indianexpress, Instagram, Ebay, Etsy, Dropbox, Github, Godaddy, Google, Icicibank, Cricbuzz, Jiosaavn, Mailchimp, Netflix, Paypal, Salesforce, Sbi, Shopify, Stackoverflow, Stackexchange, Timesofindia, Tribunnews, Twitter (X), Udemy, W3schools, Wordpress, Amazon, Atlassian, Booking, Bookmyshow, Chase, Cnn, Mediafire, Msn, Indeed, Quora, Researchgate, Roblox, Saraliitjammu, Soundcloud, Steamcommunity, Swiggy, Theguardian, Thenewyorktimes, Twitch, Walmart, Wetransfer, Zoom. \\

To generate phishing-like counterparts, we adapted our adversarial generation strategy to suit mobile layouts. The generic login form templates were redesigned using responsive CSS, ensuring they displayed appropriately on mobile screens. Furthermore, we retained the diversity strategy employed earlier by varying the form placement across different vertical sections of the page. These variations helped simulate a wide range of plausible phishing page structures typically encountered in mobile contexts.

This mobile-specific dataset contributes to a broader vision of handling layout and design differences between desktop and mobile login pages. The incorporation of mobile-optimized pages, is part of our planned future work. A preliminary discussion on this direction is provided in this Section~\ref{sec:conclusion-future}.




\subsection{Time and Space Complexity of SNM}

\begin{figure}
    \centering
    \includegraphics[width=1\linewidth]{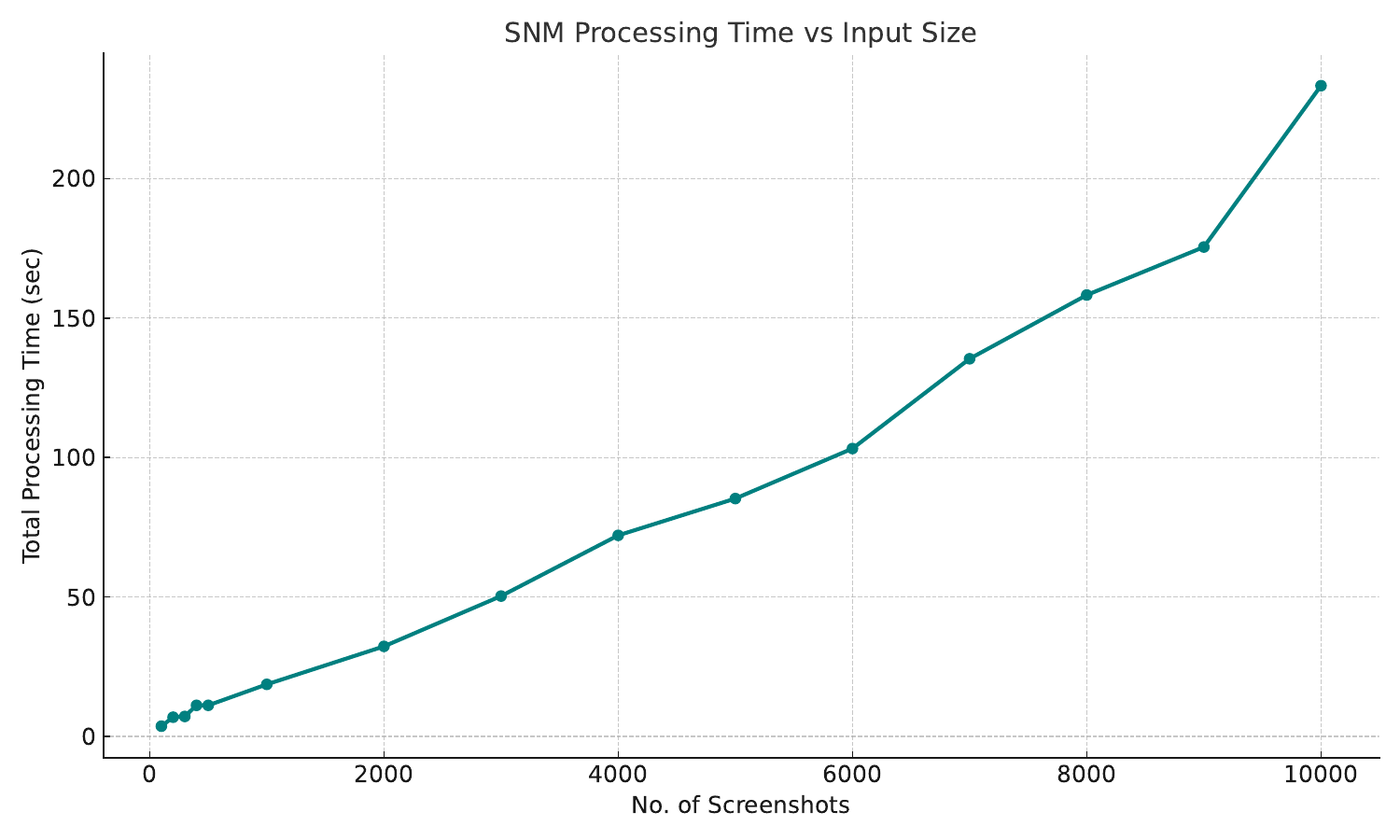}
    \caption{SNM Time Complexity}
    \label{fig:SNMtime}
\end{figure}

To evaluate the scalability of our Siamese Network Model based phishing detection system, we conducted a series of experiments measuring the processing time required to handle varying numbers of screenshots sent concurrently from the client to the server deployed with SNM processing model. The client configuration was set with MAX\_RETRIES = 20, RETRY\_DELAY = 1 second. On the server side, Flask handled requests through multi-threading, with a THREAD\_LIMIT = 10.

For each experiment, we measured two metrics:

\textbf{Individual Response Time:} Time taken for one screenshot to be processed (which consistently remained between 0 to 0.3 seconds).

\textbf{End-to-End Processing Time:} Total time taken for all screenshots to be processed, measured from the start of the client program.

The results, plotted in Figure \ref{fig:SNMtime}, show that the end-to-end processing time increases nearly linearly with the number of screenshots. For instance, processing 1,000 images took approximately 18.64 seconds, while 10,000 images took around 233.36 seconds. This indicates that our system scales in a controlled, linear fashion, without exponential degradation, even under high loads.

The main driver of space consumption in the SNM pipeline is the temporary storage of screenshots and their corresponding image embeddings. During inference, each incoming screenshot is resized to a standard input shape expected by the model and held in memory in \texttt{float32} format, contributing a fixed memory cost per image. The system also maintains a persistent in-memory store of embeddings for all known login pages. Each embedding is a fixed-size vector, resulting in a total memory requirement of $\mathcal{O}(n \cdot d)$ for $n$ stored embeddings of dimensionality $d$. A temporary buffer is used to hold embeddings for incoming query images during evaluation, adding an additional $\mathcal{O}(m \cdot d)$ memory component, where $m$ is the number of queries. Overall, the space complexity of the inference pipeline is $\mathcal{O}(n \cdot d + m \cdot d)$.

Our solution demonstrates linear scalability and efficient performance, making it highly effective as a prototype. This ensures predictable behavior even under increasing loads. As a proof of concept, it performs as well as expected for a research prototype. When implemented in a real-world system, this architecture can be further enhanced for real-time responsiveness by incorporating GPU-accelerated batch processing, asynchronous task queues. Deploying the system as a scalable microservice would also allow it to handle bursts of traffic more gracefully, making it robust enough for live phishing detection scenarios.


\subsection{Time Overhead}

We collected latency data for multiple components of our process by accessing it through the browser extension for around 180 websites.
Figure \ref{fig:latency} illustrates the average latency for each key stage in the AntiPhish browser extension architecture:

Login Page Detection Algorithm: ~0.014 s

Header Verification Algorithm: ~0.016 s

SNM Processing : ~0.11 s

Since our architecture proposes SNM to be deployed on top of PLS, so the latency caused by PLS query is expected to closely match that of SNM processing.

Each timing reflects an average of repeated measurements for consistency. 
These results indicate that the SNM Model Processing stage dominates the overall latency, which is expected due to the computationally intensive nature of embedding generation and visual comparison. In contrast, Login Page Detection and Header Verification are relatively lightweight operations and contribute minimally to the total response time.

We also measured the additional time required for websites to load when using our proposed solution, using the loading time of a standard website as the baseline. Based on the latency introduced by each component of our architecture, we evaluated two scenarios with our browser extension integrated:
\begin{itemize}
    \item When the signature is verified successfully. 
    \item When the signature is unverified and a page screenshot is sent for SNM processing.
\end{itemize}

Figure \ref{fig:time} shows the approximate time overhead introduced in each scenario compared to the baseline. When header verification succeeds, the overhead is around 9\%. If header verification fails and the Siamese model processes a screenshot, the overhead increases to approximately 43\%.

Our architecture demonstrates its feasibility for real-time deployment with acceptable latency overhead, particularly considering that SNM processing is only triggered when header verification fails — a rare event under normal conditions.
\endgroup

\begin{figure}
    \centering
    \includegraphics[width=1\linewidth]{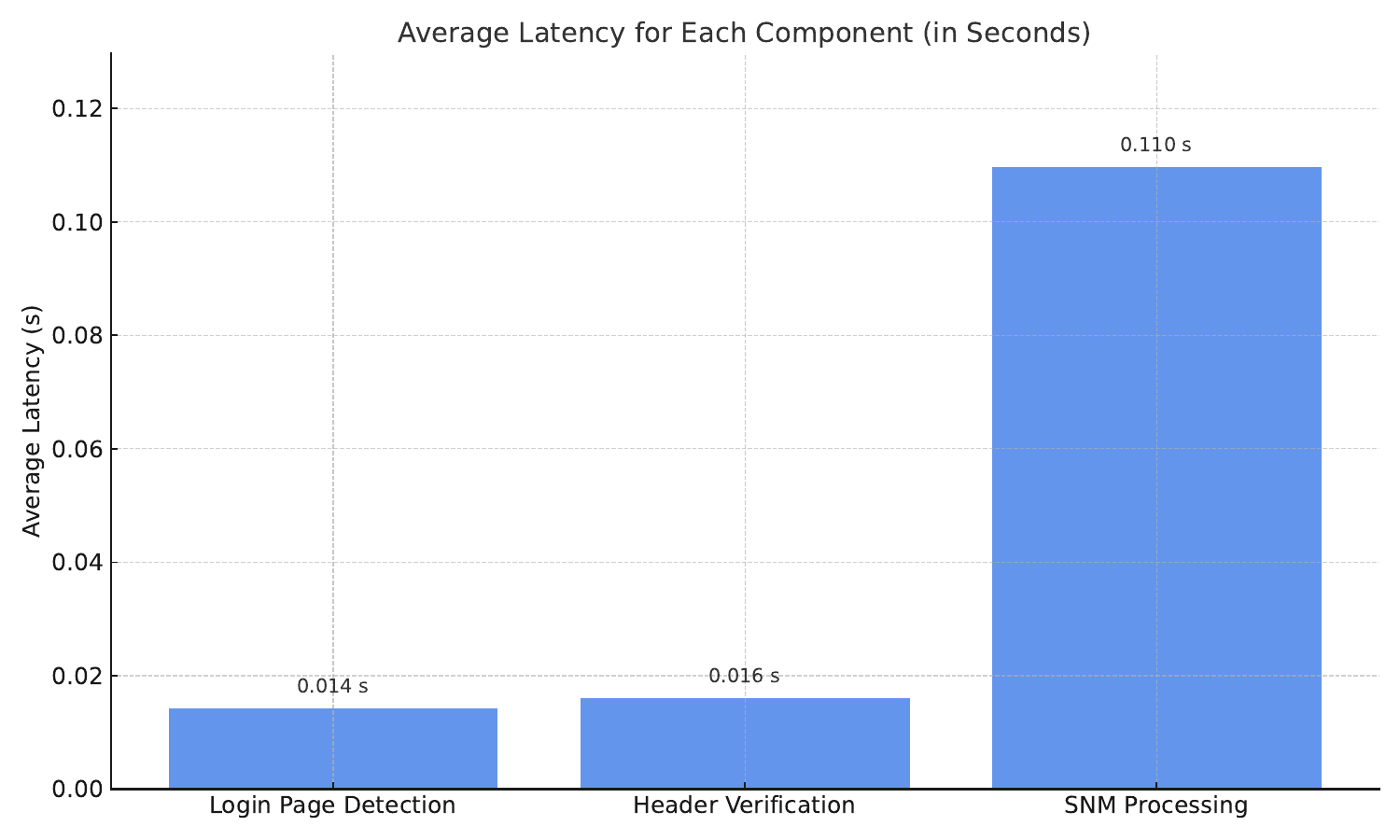}
    \caption{Latency}
    \label{fig:latency}
\end{figure}

\begin{figure}
    \centering
    \includegraphics[width=1.05\linewidth]{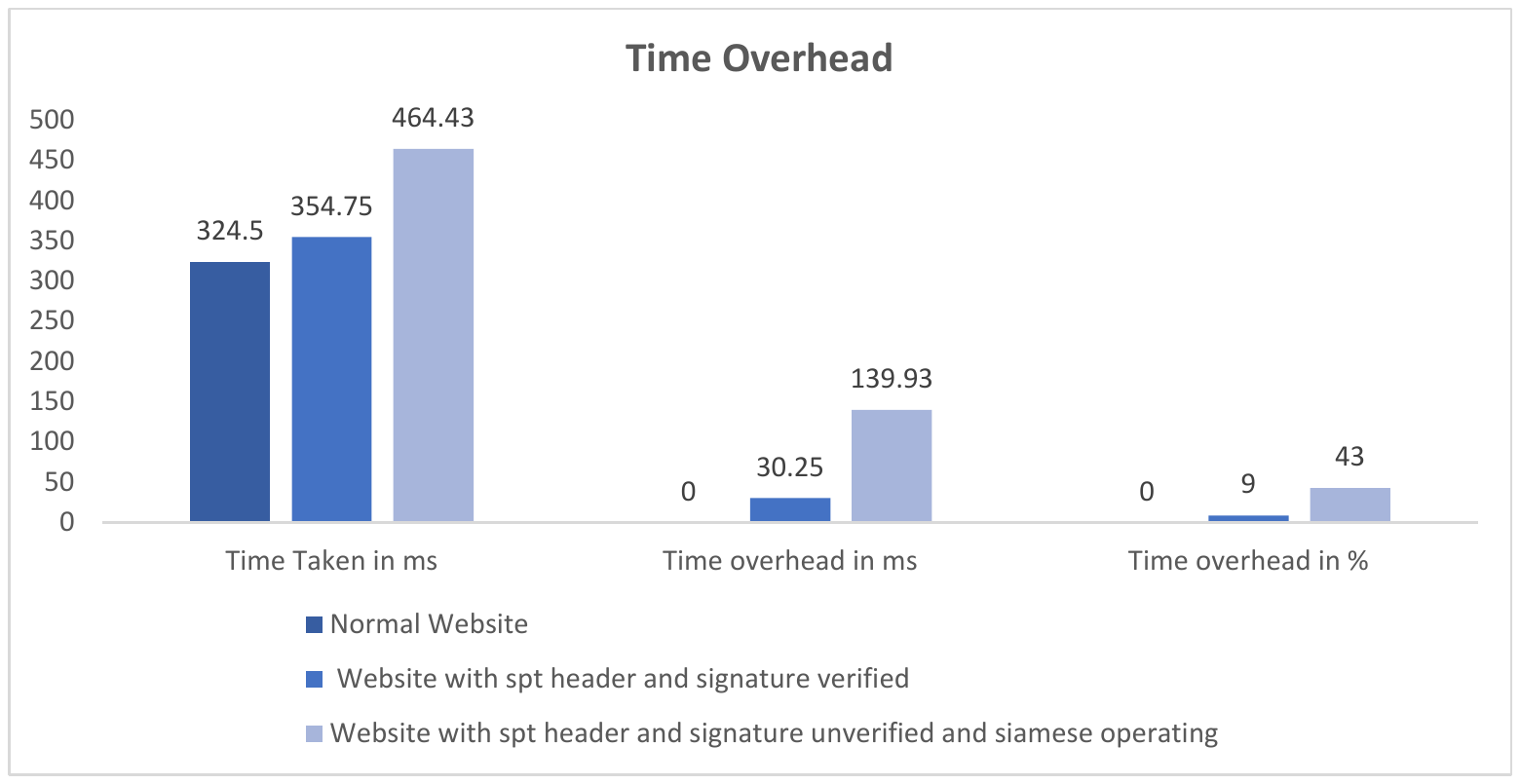}
    \caption{Time Overhead}
    \label{fig:time}
\end{figure}


\subsection{Communication Overhead}
We divide our data overhead analysis into three parts based on the different communications involved:
\begin{itemize}
    \item Client \& Server Instance1: This handles rendering the index.html page and returns the SPT\_Header, URL, and other headers.
    \item Client \& Server Instance2 : This involves communication with PLS for signature validation.
    \item Client \& Server Instance3 : This is communication required for the SNM model processing when signature validation for a page fails. 
\end{itemize}

\begin{figure}
\centering 
\includegraphics[width=1\linewidth]{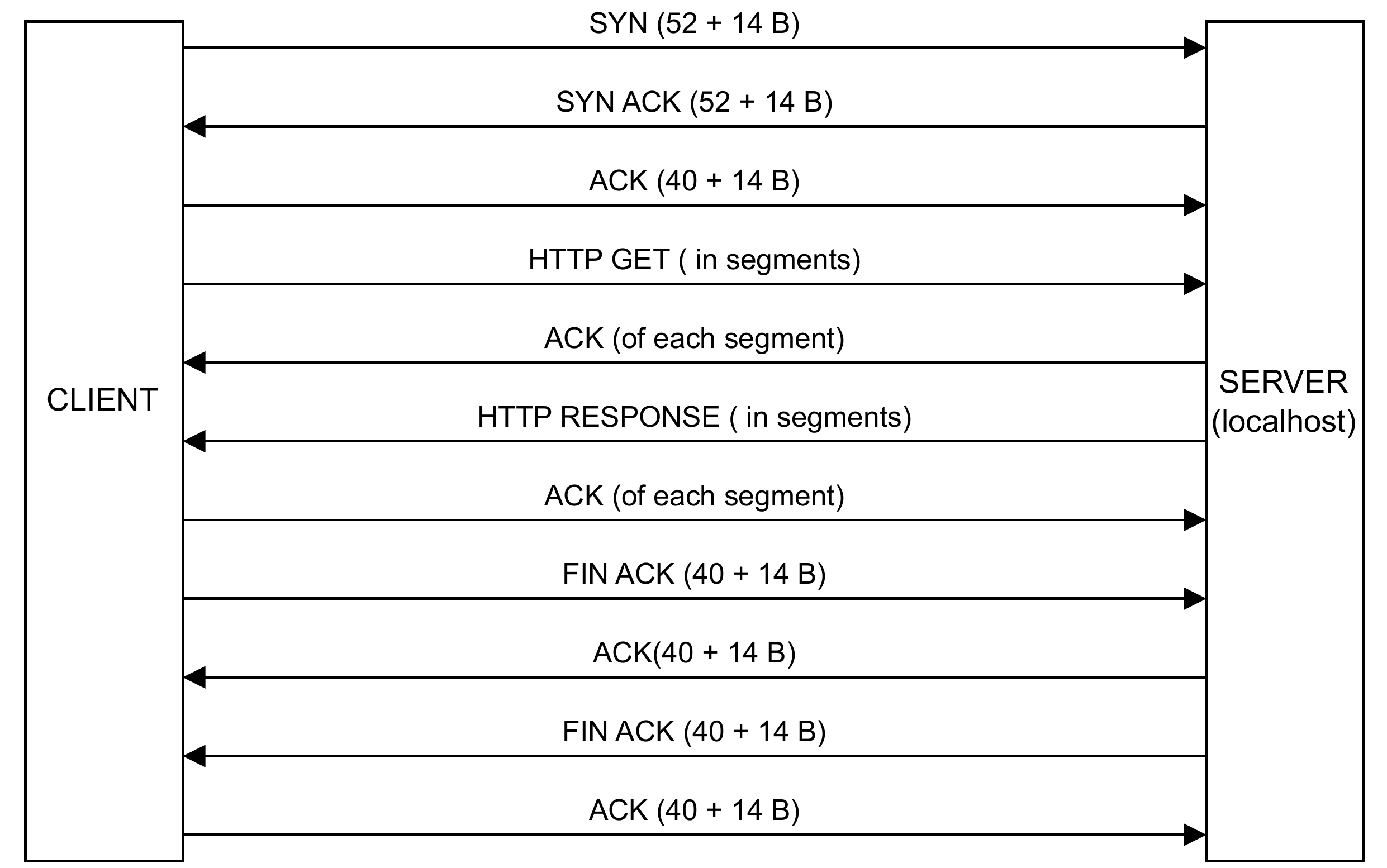}
\caption{\label{fig:17}Packets captured through Wireshark Mechanism}
\end{figure}

We captured the packets of these communications using Wireshark. Figure \ref{fig:17} shows the packets captured in one communication. The first three packets (SYN, SYN-ACK, ACK) establish the connection. The client then sends an HTTP request, which is divided into segments. The server acknowledges these segments upon receipt, processes the request, and sends a response, also segmented and acknowledged by the client. Finally, the connection is closed with the last four packets (FIN).
By summing the size of each packet, we determine the total data size transmitted in each communication, i.e. for websites for which SPT Header and signature verified, there is 112.61\% increment and for website with SPT Header and signature unverified and SNM operating this increment was 381.2\%. 

\begin{figure}
\centering 
\includegraphics[width=1\linewidth]{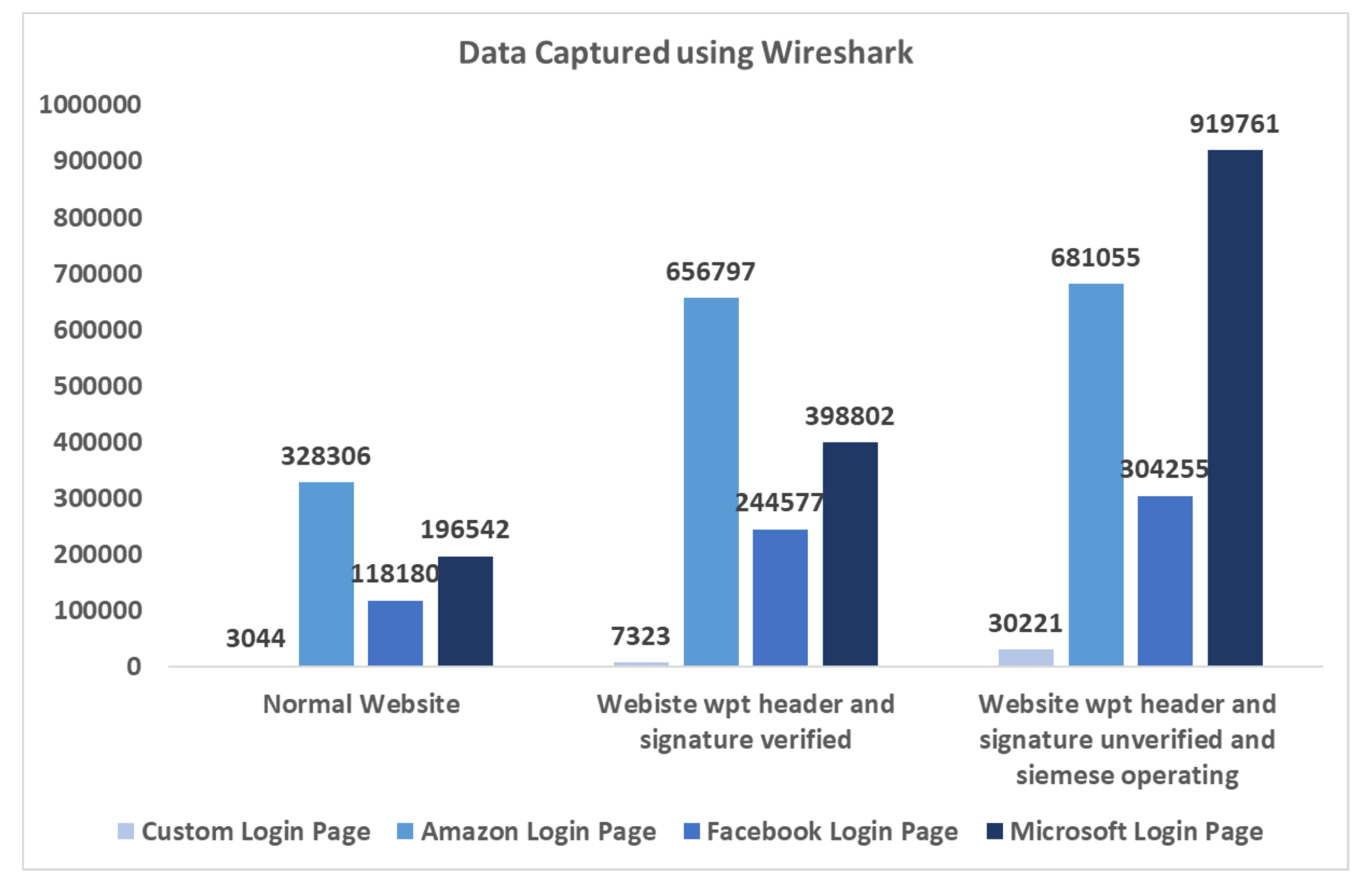}
\caption{\label{fig:18}Packets captured through Wireshark}
\end{figure}

\section{Conclusions and Future Work}
\label{sec:conclusion-future}
In this paper, we proposed an architecture to combat phishing attacks by introducing the concept of "login-page transparency." The architecture leverages PLS and SPTs to ensure that every legitimate login page is logged and verified, reducing the risk of phishing attacks. By employing a SNM for visual similarity detection, our approach effectively prevents the logging of phishing pages that visually mimic legitimate ones. 

We concentrated on detecting phishing attacks specifically within login pages, leading to the creation of a dataset of login pages of phishing and legitimate ones.
Unlike previous efforts that have addressed only certain aspects of phishing detection, we propose a comprehensive architecture that not only identifies phishing pages but also provides a foundational framework for integration into existing systems.
We have implemented Page Transparency-enabled Browsers further enhances security by ensuring that only logged pages with valid SPTs are rendered, thereby protecting users from fraudulent websites. To facilitate real-world adoption, we have developed a browser extension that integrates this architecture at a low level.

Our research indicates that the proposed architecture introduces significant overhead in both time and data transfer. Specifically, we observed an average increase of 793\% in the time required to render the worst-case scenarios. Additionally, there is a 381.2\% increase in the volume of data transmitted.

At present, we are facing challenges with the image similarity model which is SNM, a better visual similarity matching solution could be introduced to better image matching. Also, certain Hashing mechanisms may be introduced on logged pages and make these records append-only to make the system more reliable. Some mechanism could also be developed to optimize the embedding fetching overhead.
\begingroup
Though the PLS is highly trusted in the system and is a trust anchor yet in the future, users may have concerns about the privacy of their web access as the browser contacts PLS for the verification process when SPT fails. Though web request data is not stored for any historical analysis yet, in the worst case,  the IP address available to a malicious insider through a PLS can infer some short term behavioral patterns of web access for a specific IP and an unknown user behind the IP. In future to avoid this niche possibility as well the browsers can redirect the PLS request through their central servers for user anonymity. This can be also done at the user's end to have a forward proxy set for PLS access from the browser. We propose to add policy and procedures in practice to avoid any possibility of a malicious insider or a compromise, as any compromise of PLS will affect the trust of the complete system. Large-scale deployment issues and those that relate to scalability has to be looked at in depth when the solution is adopted. Our experiments and tests till date have been done at a smaller scale and in our future research, we will be study scalability and operational challenges for such a solution to operate over the Internet. This study will explore sustainability, control and scalability of the solution over a near to real world deployment.  \\

As part of our future work, we plan to maintain separate models for desktop and mobile login pages, acknowledging the structural and visual differences across device types. During page logging, the PLS will render and capture screenshots of the login interface in both desktop and mobile layouts. Each variant will then be evaluated independently using its respective model. If either the desktop or the mobile version is flagged as suspicious, the system will treat the page as a phishing page. This dual-path evaluation ensures that phishing attempts specifically designed to exploit mobile-specific layouts are not overlooked, while preserving high detection accuracy for desktop views.

\endgroup
\bibliographystyle{cas-model2-names}
\bibliography{refs}
\section{Appendix A}
\label{app:A}
\subsection{Implementation of Phase 1}
\begin{figure}[ht]
\centering
\includegraphics[width=0.98\linewidth]{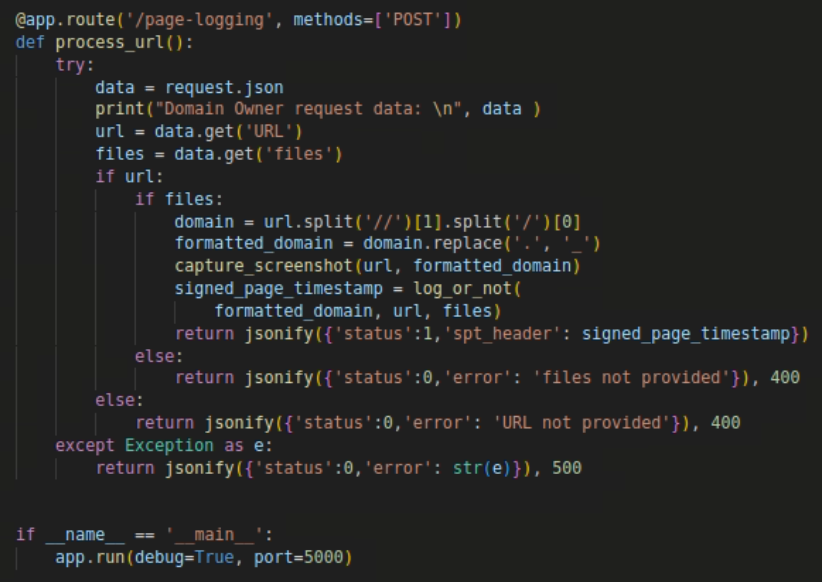}
\caption{\label{fig:PLS}Page Logging by PLS}
\end{figure}

We have set up our PLS and Domain Owner servers at different ports on localhost. For PLS, we have deployed a Flask server on the localhost network interface, listening on port 5000. This deployment ensures that the PLS is accessible for handling log requests locally. The Python code is visually represented in Figure \ref{fig:PLS}.

\begin{figure}[ht]
\centering
\includegraphics[width=0.98\linewidth]{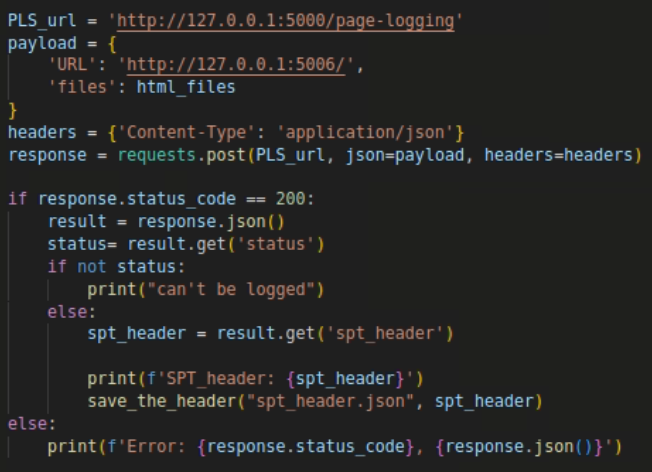}
\caption{\label{fig:domain owner client}Logging request by Domain Owner to PLS}
\end{figure}

\begin{figure*}[ht]
\centering
\includegraphics[width=0.98\linewidth]{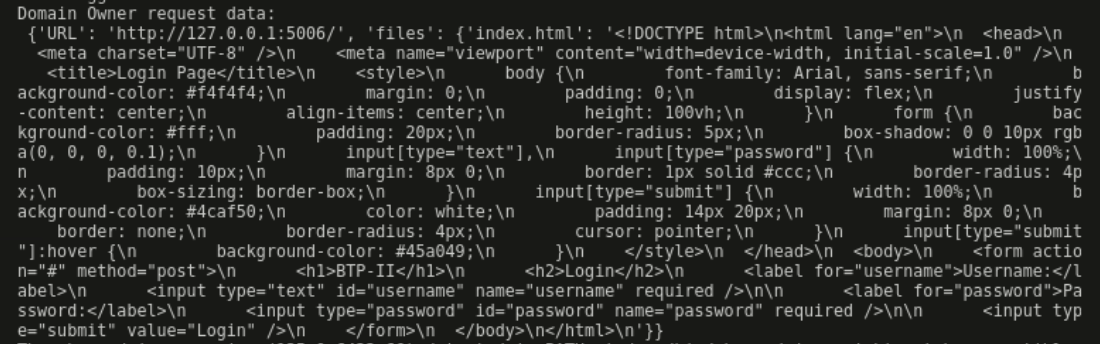}
\caption{\label{fig:domain owner request data}Domain Owner Request Payload}
\end{figure*}

The process begins with the domain owner, which initiates a request to the PLS. This step is depicted in Figure \ref{fig:domain owner client}, where the domain owner requests the URL of the PLS. The request made by the domain owner includes information, such as the URL and the HTML code of the login page. Figure \ref{fig:domain owner request data} illustrates an example of such a request, detailing how the domain owner provides both the URL and the associated HTML content of the login page.

\begin{figure}[ht]
\centering
\includegraphics[width=0.98\linewidth]{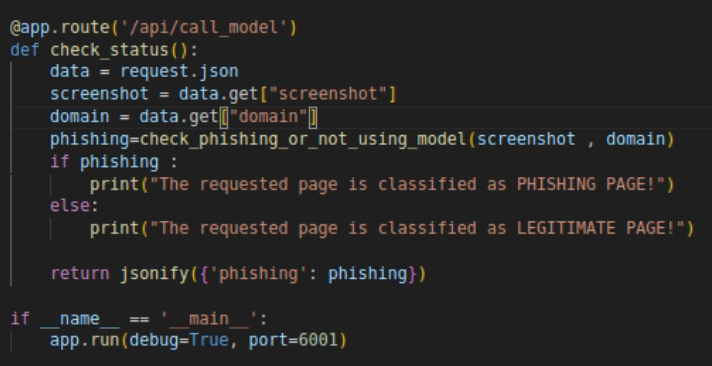}
\caption{\label{fig:Model Server}Siamese Model Server Request Handling }
\end{figure}

\begin{figure}[ht]
\centering
\includegraphics[width=0.98\linewidth]{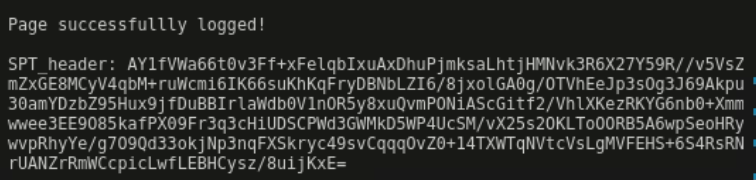}
\caption{\label{fig:Received SPT HEADER}Received SPT Header}
\end{figure}

Upon receiving request from domain owner, the PLS proceeds to interact with the model server. The model server is running on localhost at port 6001. The PLS sends a request to the model server, which includes: the domain of the requested login page and a screenshot of the login page itself.
Figure \ref{fig:Model Server} provides a comprehensive illustration of how the model server handles this request. It shows the model server receiving the data from the PLS request and calling a function to check whether the requested page is a login page or not. The requested page will be evaluated using our trained SNM. And, the result from this model will be returned to PLS as a response. Then, according to the response, PLS decides whether to log the page or not. If PLS logs the page then it returns an SPT as a response to the request of the domain owner. This response is shown in \ref{fig:Received SPT HEADER}.

\subsection{Implementation of Phase 2}
Now, we discuss implementation of "AntiPhish" browser extension.
We have these scripts running for creating a browser extension

\textbf{background.js}: Serving as the central nervous system of our extension, it orchestrates crucial functionalities. It harnesses APIs such as webRequest.onHeadersReceived to intercept and analyze web requests, onMessage for communication between different parts of the extension, and onDOMContentLoaded to react to page loads and custom functions that work together to intercept pages to be loaded on browser, extract their headers, check for spt header and validate it using signature verification. In case verification fails, sending page screenshot to SNM deployed server and finally either continue page loading or stop it from loading based on the response from SNM network.

\textbf{manifest.json}: This file serves as the blueprint for our extension's architecture. It outlines essential configurations, including the manifest version, background scripts integration, and permissions required for seamless operation. Moreover, it specifies other details like the extension's name, version, description, and icons, essential for its identification and distribution across browser platforms.

We had these different servers running on localhost to communicate with the extension.

1) \textbf{Signature validation server}: This server is called by background script with request data(headers of the page loaded on the browser, URL, HTML content of the page). It then verifies the spt header and returns the response as a signature verified or not. Algorithm 3 can be referred to as the pseudo-code.

2) \textbf{Model deployed server}: An API call is made from the background script with screenshot of current page to be sent to model deployed server in case the signature validation server responds with signature unverified. The screenshot is sent to the SNM deployed server, which compares screenshots embeddings with logged pages embeddings and responds with a phishing/no phishing page response accordingly.  Algorithm \ref{algo: Model Server} can be referred for the pseudo code of model deployed server.

Pseudo code for background script is shown in Algorithm \ref{algo: Extesion Working}.

The process begins when user searches for some website on their browser.
Our extension acts as an inteface for conducting all the checks before a page gets loaded on the browser, 
Since our main concern is only login pages, when a user loads a page on the browser, our extension, using the algorithm \ref{algo:LPDA} explained in section \ref{sec:PRP}, checks for login pages.  If it's not a login page, the page is directly loaded. Otherwise, it on receiving the headers using Web Request API, sends the headers along with the html content to the signature validation server that conducts verification.
If the server responds with a signature-verified response. the extension does not hamper the page loading. If the server responds signature is unverified, it means according to the records present in the PLS, the current page cannot be verified and we need to check whether the visual appearance of the page matches with any other page. For that, our phishing detection model comes into the picture, a screenshot of the page is taken(explained below) by the extension and is sent to the model deployed server as shown in Algorithm \ref{algo:SVA}. The model checks the visual similarity of the page with all logged pages and make a decision wheather page is phishing or not. With a phishing response, the extension(Algorithm \ref{alg:SPT} ) stops the loading of the page with warnings that the page is phishing. If no phishing is found by model, the browser does not interfere with the loading of the page.

We tried these 5 cases, that cover all possible scenerios. The diagrams corresponding to each of the cases can be found. These show the transition of the page from left to right as our extension interferes with the rendering of the page.

\begin{figure*}[ht]
\centering
\includegraphics[width=0.98\linewidth]{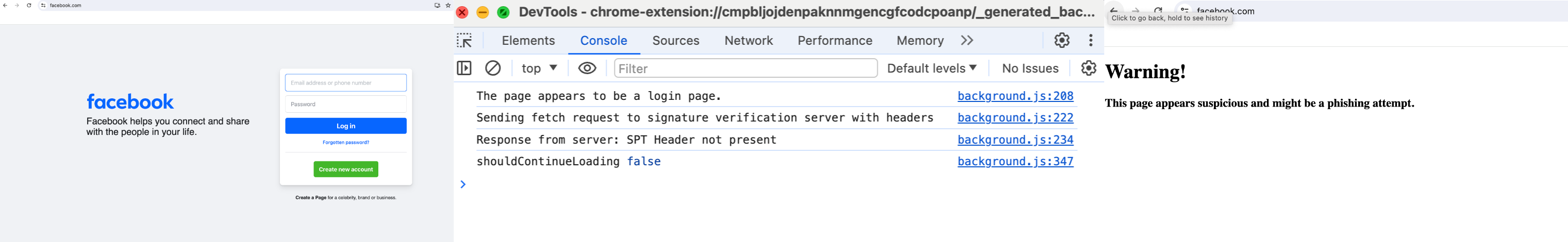}
\caption{\label{fig:LoginNoPtHeader}Login Page with no SPT Header present}
\end{figure*}

1) Login Page(No pt-header present) Figure \ref{fig:LoginNoPtHeader}

\begin{figure*}[ht]
\centering
\includegraphics[width=0.98\linewidth]{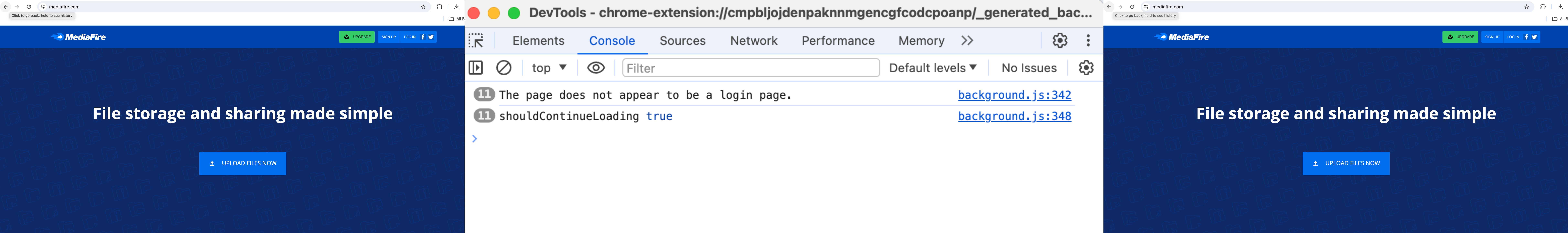}
\caption{\label{fig:Non Login Page}Non Login Page}
\end{figure*}

2) Non Login Page(No pt-header present) Figure \ref{fig:Non Login Page}

\begin{figure*}[ht]
\centering
\includegraphics[width=0.98\linewidth]{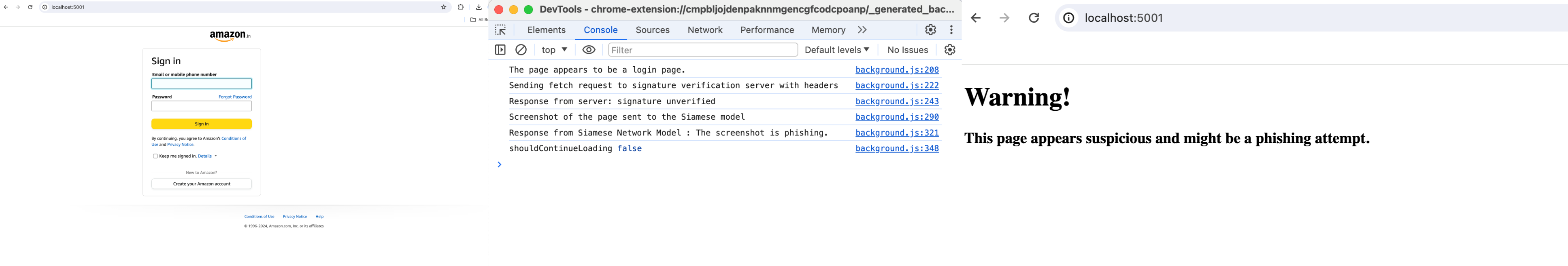}
\caption{\label{fig:Phishing}Signature Unverified and declared Phishing}
\end{figure*}

3) Login Page(pt-header present, signature unverified, model detects as phishing) Figure \ref{fig:Phishing}

\begin{figure*}[th]
\centering
\includegraphics[width=0.98\linewidth]{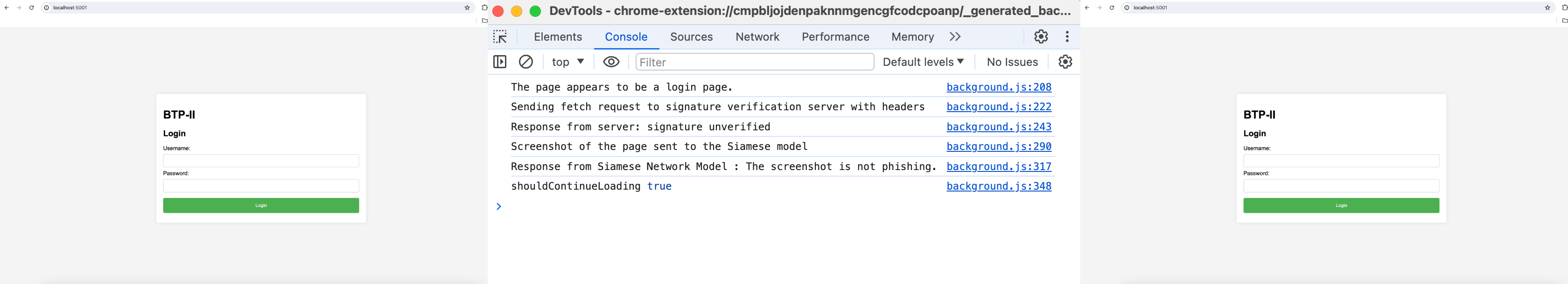}
\caption{\label{fig:notPhishing}Signature Unverified and declared Non Phishing}
\end{figure*}

4) Login Page(pt-header present, signature unverified, model detects as non-phishing) Figure \ref{fig:notPhishing}

\begin{figure*}[h]
\centering
\includegraphics[width=0.98\linewidth]{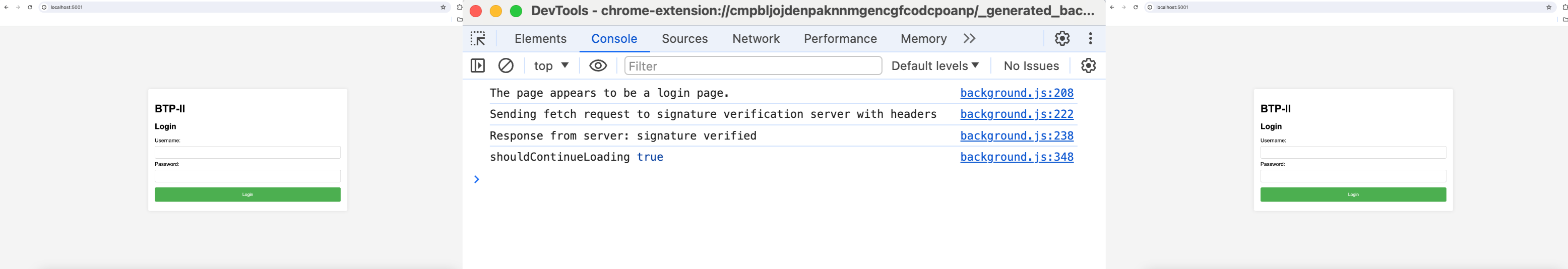}
\caption{\label{fig:Header Verified}Signature Verified}
\end{figure*}

5) Login page(pt-header present, signature verified) Figure \ref{fig:Header Verified}. \\
\\ \textbf{How we take screenshots of webpages from URLs}\hfill\\
We can install a WebDriver for our preferred browser, if not installed by default ensuring the WebDriver version matches our browser version.
Chrome WebDriver, specifically, is a WebDriver implementation for the Chrome web browser. It allows us to control and automate Chrome browser actions programmatically using Selenium WebDriver's API. With Chrome WebDriver, we can automate tasks in Chrome such as opening webpages, filling out forms, clicking buttons, and taking screenshots.
In our case, WebDriver was already installed for Chrome Browser by default. We just installed selenium, a Python package, and used it in our code.







-
\end{document}